\documentclass[prd,nofootinbib,amsmath,amssymb,superscriptaddress,preprintnumbers,twocolumn]{revtex4-1}

\usepackage[dvipdfmx]{graphicx}
\usepackage[colorlinks=true,link color=blue,cite color=blue,urlcolor=black]{hyperref}

\begin{document}
\preprint{ICRR-Report-681-2014-7, RESCEU-10/14}
\title{No quasi-stable scalaron lump forms after $R^2$ inflation}

\author{Naoyuki Takeda}\email{takedan@icrr.u-tokyo.ac.jp}
\affiliation{Institute for Cosmic Ray Research, University of Tokyo, Kashiwa, Chiba 277-8582, Japan}
\author{Yuki Watanabe}\email{watanabe@resceu.s.u-tokyo.ac.jp}
\affiliation{Research Center for the Early Universe, University of Tokyo, Tokyo 113-0033, Japan}

\date{\today}

\begin{abstract}
In the Einstein frame picture of Starobinky's $R^2$ inflation model, cosmic inflation is driven by a slowly rolling inflaton field, called scalaron, and followed by a coherently oscillating scalaron phase.
Since the scalaron oscillates excessively many times in its potential, which has a quadratic minimum and is a little shallower than quadratic on the positive side, it may fragment into long-living localized objects, called oscillons or I-balls, due to nonlinear growth of fluctuations before reheating of the universe.
We show that while parametric self-resonances amplify scalaron fluctuations in the Minkowski background, the growth cannot overcome the decay due to expansion in the Friedmann background after $R^2$ inflation.
By taking into account back-reaction from the metric of spacetime, modes that are larger than a critical scale are indeed amplified and become non-decaying. However, those non-decaying modes are not growing enough to form spatially localized lumps of the scalaron.
Thus, reheating processes are unaltered by oscillons/I-balls and they proceed through perturbative decay of the scalaron as studied in the original work.
\end{abstract}


\maketitle

\section{Introduction}
The $R^2$ inflation \cite{Starobinsky:1980te,Starobinsky:1981vz} is one of the oldest models of inflation \cite{Kazanas:1980tx} proposed even before puzzles of the hot big bang cosmology, such as the horizon, flatness, homogeneity, isotropy, entropy, and monopole problems, were claimed to be solved by the quasi-exponential expansion of the universe \cite{Sato:1980yn, Guth:1980zm, Linde:1981mu, Albrecht:1982wi}. 
Soon after the model was announced by Starobinsky, Mukhanov and Chibisov \cite{Mukhanov:1981xt,Chibisov:1982nx} realized that cosmological fluctuations resulting from quantum fluctuations were generated to seed the almost scale-invariant initial density fluctuations (with a logarithmic dependence on wave number) to form cosmic structures like galaxies and galactic clusters.
The $R^2$ inflation is unique in the sense that the quasi-DeSitter phase is achieved only by gravitational degrees of freedom; a homogeneous condensate of scalar gravitons (dubbed \textit{scalarons}) slowly evolves to an oscillating phase and perturbatively decays into relativistic particles via production of conformally non-invariant particles that exist in the universe \cite{Starobinsky:1981vz,Vilenkin:1985md}. 

Observationally, anisotropy of the Cosmic Microwave Background (CMB) is measured in the full sky with multiple frequency bands and high precision by satellite missions, such as WMAP \cite{Hinshaw:2012aka} and Planck \cite{Ade:2013uln}, to test inflationary predictions.
Remarkably, Mukhanov and Chibisov~\cite{Mukhanov:1981xt} predicted that the primordial spectrum of curvature fluctuations for $k\ll aH$ is given by (see Eq.~(9) of \cite{Mukhanov:1981xt} and Appendix of \cite{Hinshaw:2012aka} for a review)
\begin{align}
{\cal P}_{\zeta}(k)\sim \frac{M^2}{M_p^2} \left(1+\frac12 \ln{\frac{aH}{k}}\right)^2,
\end{align}
where $M$ is the scalaron mass and $M_p\equiv 1/\sqrt{8\pi G}$ the reduced Planck mass. It implies a slightly red spectrum with
\begin{align}
n_s-1= \left.\frac{d \ln{\cal P}_{\zeta}(k)}{d\ln{k}}\right|_{k=k_*}=-\frac{2}{N_*+2},
\end{align}
where $N_*$ is the number of e-folds at the CMB scale.
With $N_* = 50$, for instance, $n_s \approx 0.96$, which is in good agreement with \cite{Hinshaw:2012aka,Ade:2013uln}.

Even though high-precision observations have excluded many classes of inflationary models, some physics of inflation remain to be unveiled so as to further constrain the observationally qualified models, including the $R^2$ inflation.
Especially, the physics of reheating after inflation is not fully understood since fluctuations of matter fields or the inflaton itself would be enhanced by parametric resonances due to the oscillation of the 
inflaton, called \textit{preheating} \cite{Shtanov:1994ce,Kofman:1994rk} (also see \cite{Dolgov:1989us} for earlier works). 
Understanding of the reheating epoch is getting more important as the precise value of $N_*$ depends on duration of this epoch and results in precise predictions of inflationary models.
The nonlinear evolution of matter and/or inflaton fluctuations due to preheating, if it happens, changes the dynamics of the inflaton decay~\cite{Kofman:1994rk} and/or produces observable signals, such as gravitational waves on small scales \cite{Khlebnikov:1997di, Bassett:1997ke}. 

It has been pointed out that the non-perturbative effect enhances the inflaton fluctuations during the post-inflationary oscillation phase, such that the coherent inflaton field fragments into massive, long-lived, localized objects (often called oscillons~\cite{Copeland:1995fq} or I-balls~\cite{Kasuya:2002zs}) if the inflationary potential is shallower than quadratic away from the minimum~\cite{Amin:2010xe} (also see \cite{Greene:1998pb, McDonald:2001iv} in different contexts).
The shallow potentials include Starobinsky's $R^2$ inflation model in the Einstein frame. 
The oscillons/I-balls are the quasi-stable, spherical, soliton-like solution of non-linear scalar field excitations whose pressure from the gradient term is balanced with the attractive force from the shallower potential; thus once they are formed, their decays are limited only on the surface with constant energy density decoupled from the Hubble expansion.
The decay rate of the inflaton is therefore significantly altered from the standard perturbative decay~\cite{McDonald:2001iv,Hertzberg:2010yz,Kawasaki:2013awa}. 

In previous works~\cite{Amin:2010xe} the criteria of formation of oscillons/I-balls were obtained. 
The Floquet theory indicates that each k-mode of inflaton fluctuations has an exponential instability $\delta\phi_k(t) = p_+(t)e^{\mu_k Mt}+p_-(t)e^{-\mu_k Mt}$ within the set of resonance bands during preheating, where $M$ is the inflaton mass, $p_{\pm}(t)$ are periodic functions and $\mu_k$ is the Floquet exponent. If $\mu_k M$ is real and larger than the Hubble rate $H_f$ at the end of inflation, the inhomogeneous $k$-mode can grow exponentially; resonance generates large over-densities that may fragment into oscillons/I-balls.
Especially the fourth Ref.~of~\cite{Amin:2010xe} revealed that strong resonance with $[\mu_k M/H]_{\rm max} \gtrsim 10$ is both necessary and sufficient for prompt, copious formation of oscillons/I-balls by performing numerical simulations with a fairly generic class of inflaton potentials.

In this paper, we study the dynamics of the inflaton field (scalaron) during the post-inflationary oscillation epoch after $R^2$ inflation~\cite{Starobinsky:1980te}. 
This inflation model is highly motivated since it is parametrized only by one parameter and its prediction of the spectral index and the tensor-to-scalar ratio is located at the central value of observational results by WMAP~\cite{Hinshaw:2012aka} and Planck~\cite{Ade:2013uln}.\footnote{The BICEP2 collaboration~\cite{Ade:2014xna} has recently announced the detection of B-mode polarization at degree angular scales and provided the tensor-to-scalar ratio $r=0.16^{+0.06}_{-0.05}$ with the dust removal, which is in tension with the full-sky multiple-band temperature result of Planck~\cite{Ade:2013uln}. Since BICEP2 could reject the dust component only at 2.2$\sigma$, their result should be confirmed by other experiments~\cite{Mortonson:2014bja}.}
In addition, since the identity of the inflaton is homogeneous condensate of scalar gravitons, it universally interacts with elementary particles in the standard model (and beyond), only with gravitational strength; thus, the perturbative decay processes can be computed with the known parameters. 
However, the nonlinear evolution of the scalaron fluctuations themselves is yet fully understood.
The potential of Starobinsky's inflation is shallower than quadratic away from the minimum on the positive side; thereby, the scalaron fluctuations could be amplified during the reheating epoch and fragment into quasi-stable lumps (oscillons/I-balls). 
In this case, the perturbative analysis of reheating after $R^2$ inflation is significantly altered. Therefore, we will clarify this issue.

The organization of this paper is the following. 
First, we review Starobinsky's model of inflation and its predictions in Sec.~\ref{Sec_model}.
Second, we study the enhancement of the scalaron fluctuations in a static Minkowski spacetime ($H=0$) in Sec.~\ref{Sec_insH0}. 
We then study the enhancement in an expanding Friedmann spacetime in Sec.~\ref{Sec_insHn0}.
Finally, we make concluding remarks in Sec.~\ref{Sec_CONCLUSION}. 
We work with the metric signature $(-,+,+,+)$, $R^{\alpha}{}_{\mu\beta\nu}=\partial_{\beta}\Gamma^{\alpha}{}_{\mu\nu}+\cdots$ and $R_{\mu\nu}=R^{\alpha}{}_{\mu\alpha\nu}$ for convention.

\section{Summary of Starobinsky's $R^2$-inflation}
\label{Sec_model}
We consider Starobinsky's model~\cite{Starobinsky:1980te} whose action is given by
\begin{align}
\label{eq_model_1}
S &= \int d^4x\sqrt{-g}\frac{M_p^2}{2}F(R) +\int d^4x {\cal L}_{\rm matt}[g_{\mu\nu},\chi,\psi ,A_{\mu}],\nonumber\\
&F(R)= R+\frac{R^2}{6M^2}.
\end{align}
It is well known that the above $F(R)$ gravity~(\ref{eq_model_1}) is conformally equivalent to the Einstein gravity with a scalar field minimally coupled to gravity \cite{Bicknell:1974,Maeda:1987xf,Mukhanov:1989rq,Jakubiec:1988ef}. 
Let us first define a scalar field $f$ as
\begin{align}
f = f(R) = F'(R).
\end{align} 
Since $f(R)=1+R/(3M^2)$ is invertible, we can invert the functional to get
\begin{align}
R = R(f).
\end{align}
The Legendre transform of the gravitational Lagrangian is then given by
\begin{align}
F^*(f)&=F[R(f)],\quad
U(f)=fR(f) - F^*(f),\\
{\cal L}_{\rm grav}&=\frac{M_p^2}{2}\sqrt{-g}[fR -U(f)].
\end{align}
As the scalar $f$ is a propagating degree of freedom, we can add a ``gauge-fixing" term to define its propagator:
\begin{align}
{\cal L}_{\rm grav}&=\frac{M_p^2}{2}\sqrt{-g}[fR -U(f) - \gamma(\partial f)^2],
\end{align}
where $\gamma$ is a scalar function and usually positive definite from unitarity. Note however that a propagating degree of freedom $\phi=M_p\gamma^{1/2}f$ would be a non-ghost even when $\gamma$ is negative in the presence of the non-minimal gravitational coupling $f(\phi)R$.
In this way, we can regard the $R^2$ gravity as Brans-Dicke type $f(\phi)R$ gravity theories.

Let us next rescale the metric as 
\begin{align}
\hat{g}_{\mu\nu}&= fg_{\mu\nu},\\
{\cal L}_{\rm grav}& = \frac{M_p^2}{2}\sqrt{-\hat{g}}\left[\hat{R} -\frac32\hat{g}^{\mu\nu}\partial_{\mu}(\ln{f})\partial_{\nu}(\ln{f})\right.\nonumber\\
&\quad \left. -\frac{\gamma}{f}(\hat{\partial}f)^2 -\frac{U(f)}{f^2} \right],\label{eq:lag_grav_f}
\end{align}
where $\sqrt{-g}=f^{-2}\sqrt{-\hat{g}}$ and $R=f[\hat{R}+3\hat\Box(\ln{f})-\frac32\hat{g}^{\mu\nu}\partial_{\mu}(\ln{f})\partial_{\nu}(\ln{f})]$ have been used and a total derivative term has been set to zero.
Since $df=f'(\phi)d\phi$ is integrable, we redefine the scalar field $f$ to be canonically normalized as
\begin{align}
\phi = M_p\int d\tilde{\phi}f'(\tilde\phi)\left(\frac{3}{2f^2} + \frac{\gamma(\tilde\phi)}{f}\right)^{1/2},
\end{align}
which makes sense only if the integrand inside the square root is positive definite.
In fact, 
one can choose $\gamma=0$ without having a problem of strong coupling, thanks to the presence of the second term in Eq.~(\ref{eq:lag_grav_f}). In this case, the integration is easily carried out and we get
\begin{align}
f &= 1+\frac{R}{3M^2}= {\rm e}^{\sqrt{\frac23}\frac{\phi}{M_p}},\label{eq:f-phi}\\
V(\phi)&=\frac{M_p^2U(f)}{2f^2}=\frac{3M^2M_p^2(f-1)^2}{4f^2}\nonumber \\
&= \frac{3M^2M_p^2}{4}\left(1-{\rm e}^{-\sqrt{\frac23}\frac{\phi}{M_p}}\right)^2.\label{eq:pot}
\end{align}
Then the gravitational Lagrangian reads \cite{Maeda:1987xf,Mukhanov:1989rq}
\begin{equation}
{\cal L}_{\rm grav}=\sqrt{-g}\left[\frac{M_p^2}{2}R -\frac{1}{2}\partial^{\mu}\phi\partial_{\mu}\phi
-V(\phi) \right],
\end{equation}
where
we have redefined $R$ with the rescaled metric and removed carets  ( $\hat{}$ ) on variables. We call this conformally (Weyl) rescaled frame the Einstein frame, where $\phi$ is the canonically normalized scalar field that drives inflation in the early universe (dubbed inflaton or scalaron). 

In the Einstein frame picture, inflation takes place during slow rolling of the scalaron on the flat part of its potential $V(\phi)$ [Eq.~(\ref{eq:pot})].
Therefore, we can use standard formulas on potential-driven slow roll inflation in the literature (see, e.g., \cite{Liddle:2000cg,Ade:2013uln} for reviews).
The power spectra of primordial curvature and gravitational wave perturbations ($\zeta_k$ and $\gamma_k^{+,\times}$, respectively) on large scales are given by
\begin{align}
{\cal P}_{\zeta}(k)&= \frac{k^3}{2\pi^2}|\zeta_k|^2 \simeq \frac{H^2}{8\pi^2\epsilon M_p^2} \simeq \frac{V}{24\pi^2\epsilon_VM_p^4},\label{eq:Pk_zeta}\\
{\cal P}_{\gamma}(k)&= \frac{k^3}{\pi^2}|\gamma_k|^2\simeq \frac{2H^2}{\pi^2M_p^2}\simeq \frac{2V}{3\pi^2M_p^4},\label{eq:Pk_gamma}
\end{align}
where all quantities are evaluated at the CMB scale $k=k_*$ and slow roll parameters are defined as
\begin{align}
\epsilon = -\frac{\dot H}{H^2},\quad
\epsilon_V= \frac{V'^2M_p^2}{2V}.\label{eq:epsilon}
\end{align}
From Eqs.~(\ref{eq:pot}), (\ref{eq:Pk_zeta}) and (\ref{eq:epsilon}), we get
\begin{align}
{\cal P}_{\zeta}(k_*)\simeq \frac{3M^2}{128\pi^2M_p^2}{\rm e}^{2\sqrt{\frac23}\frac{\phi_*}{M_p}} \simeq \frac{N_*^2M^2}{24\pi^2M_p^2},
\end{align}
where the number of e-folds is given by
\begin{align}
N_* = -\int^*_f dt H \simeq \frac34 {\rm e}^{\sqrt{\frac23} \frac{\phi_*}{M_p}}.
\end{align}
The unique parameter, scalaron mass $M$, is thus fixed by the COBE-WMAP normalization of the amplitude of curvature perturbations as \cite{Mukhanov:1987pv,Mukhanov:1989rq,Hwang:2001pu,Faulkner:2006ub}
\begin{align} 
M &\simeq 10^{-5}M_p\frac{4\pi\sqrt{30}}{N_*}\left(\frac{{\cal P}_{\zeta}(k_*)}{2\times 10^{-9}}\right)^{1/2} \\
&\sim 10^{-5}~M_p\sim10^{27}{\rm cm}^{-1}\sim10^{51}{\rm Mpc}^{-1},\nonumber
\end{align} 
which is roughly the physical size of the Hubble horizon at the end of inflation.

The primordial amplitude of  gravitational waves is characterized by the ratio between Eqs.~(\ref{eq:Pk_gamma}) and (\ref{eq:Pk_zeta}):
\begin{align}
r = \frac{{\cal P}_{\gamma}(k)}{{\cal P}_{\zeta}(k)} \simeq 16\epsilon \simeq \frac{12}{N_*^2}.
\end{align}

Scale dependences of the primordial spectra are given by 
\begin{align}
n_s-1 &= \frac{ d\ln{ {\cal P}_{\zeta}(k) } }{d\ln{k}} \simeq -6\epsilon_V + 2\eta_V \simeq -\frac{2}{N_*},\\
n_t &= \frac{d\ln{{\cal P}_{\gamma}(k)}}{d\ln{k}} \simeq -2\epsilon_V \simeq -\frac{3}{2N_*^2},  \\
\frac{d n_s}{d\ln{k}} &\simeq 16\epsilon_V\eta_V -24\epsilon_V^2-2\xi_V^2 \simeq -\frac{2}{N_*^2},\\
\frac{d n_t}{d\ln{k}} &\simeq 4\epsilon_V\eta_V -8\epsilon_V^2 \simeq -\frac{3}{N_*^3},
\end{align}
where we have used $\epsilon_V \simeq 3/(4N_*^2)$, $\eta_V = V''M_p^2/V\simeq -1/N_*$ and $\xi_V^2 = V'V'''M_p^4/V^2\simeq 1/N_*^2$.

A precise value of $N_*$ depends on particle contents of the universe and how they couple to the inflaton during reheating.
How does reheating take place after the $R^2$-inflation? 
The matter sector is assumed to be conformally coupled to gravity in the original work~\cite{Starobinsky:1980te,Starobinsky:1981vz}, where he estimated the gravitational decay rate of the scalaron by using Bogoliubov's method in the Jordan frame (see also \cite{Vilenkin:1985md, Arbuzova:2011fu}). 
In this frame, the Ricci scalar becomes dynamical, contrary to general relativity, and starts oscillating after inflation.
It is similar to a dust-dominated phase.

In the Einstein frame picture, we can do the equivalent analysis. 
Expanding the scalaron potential~(\ref{eq:pot}) around the origin, we get $V(\phi) \simeq M^2\phi^2/2+\cdots$ for $\phi \lesssim M_p$.  
Thus we can interpret an oscillating homogeneous field $\phi$ as a condensate of massive scalar particles (scalarons) with zero momenta and mass $M$.
When the metric is rescaled, interaction between the scalaron and matter sector is semi-classically \cite{Watanabe:2006ku} and quantum mechanically \cite{Watanabe:2010vy}
 induced as \begin{align}
\frac{{\cal L}_{\rm matt}}{\sqrt{-\hat{g}}}=&  -\hat{g}^{\mu\nu}({\cal D}_{\mu}\hat\chi)^*{\cal D}_{\nu}\hat\chi
 -\lambda_\chi(\hat\chi^*\hat\chi)^2 -\frac{m_{\chi}^2}{f}\hat{\chi}^*\hat\chi \nonumber\\
& -\hat{\bar\psi}\left[ \hat{e}^{\mu}{}_{\alpha}\gamma^{\alpha}(\partial_{\mu}-\hat{\Gamma}_{\mu}-ig\hat{A}_{\mu})+f^{-1/2}m_{\psi}\right]\hat\psi \nonumber\\
& -\frac14 \hat{F}^{\mu\nu}\hat{F}_{\mu\nu}
+\frac{\beta_h(g)}{2g}(\ln{f})\hat{F}^{\mu\nu}\hat{F}_{\mu\nu},\label{eq:lag_matt}
\end{align}
where the standard model is symbolically treated as the matter sector in which fields of spin-0 ($\chi$), spin-1/2 ($\psi$) and spin-1 ($A_{\mu}$) are rescaled as
\begin{align}
\hat\chi &= f^{-1/2}\chi,\quad
\hat\psi = f^{-3/4}\psi, \\
\hat{A}_{\mu} &= A_{\mu},\quad
\hat{A}^{\mu}= f^{-1}A^{\mu},
\end{align}
respectively, and the covariant derivative for scalars is defined as
\begin{align}
{\cal D}_{\mu}\hat\chi = \partial_{\mu}\hat{\chi}+\hat\chi\partial_{\mu}(\ln{f^{1/2}}) - ig\hat{A}_{\mu}\hat\chi.\label{eq:cov_scalar}
\end{align}
The spin connection is conformally invariant: $\hat{\Gamma}_{\mu}=\Gamma_{\mu}$ (see footnote 4 of \cite{Watanabe:2010vy}). The gauge coupling constant is denoted by $g$ and its running is associated with the beta function from heavy intermediate particles $\beta_h(g)$.
Inserting $\ln{f}= \sqrt{2/3}(\phi/M_p)$ into Eq.~(\ref{eq:cov_scalar})  while expanding Eq.~(\ref{eq:f-phi}) as
\begin{align}
f = 1 + \sqrt{\frac23}\frac{\phi}{M_p} + \frac13 \left(\frac{\phi}{M_p}\right)^2+ \cdots,
\end{align} 
we get order by order expansion of the interaction Lagrangian with respect to $\phi$.
The scalaron $\phi$ can decay into the matter sector via trilinear interactions \cite{Watanabe:2010vy}:
\begin{align}
\frac{{\cal L}_{\rm 3leg}}{\sqrt{-g}} =& 
\frac{-1}{\sqrt6 M_p}\chi\partial^{\mu}\chi^*\partial_{\mu}\phi
-\frac{1}{\sqrt6 M_p}\chi^*\partial^{\mu}\chi\partial_{\mu}\phi \nonumber\\
&+ \frac{2m_{\chi}^2}{\sqrt{6}M_p}\phi\chi^*\chi 
+\frac{m_\psi^2}{\sqrt6 M_p}\phi\bar\psi\psi \quad \nonumber\\
&
+\frac{\beta_h(g)}{2\sqrt6 gM_p}\phi{F}^{\mu\nu}{F}_{\mu\nu}\nonumber\\
=&\frac{2}{\sqrt6 M_p}\phi\partial^{\mu}\chi^*\partial_{\mu}\chi
+ \frac{4m_{\chi}^2}{\sqrt{6}M_p}\phi\chi^*\chi
+\frac{m_\psi^2}{\sqrt6 M_p}\phi\bar\psi\psi \nonumber\\ 
&+\frac{\beta_h(g)}{2\sqrt6 gM_p}\phi{F}^{\mu\nu}{F}_{\mu\nu},
\end{align}
where we have integrated by parts, used equations of motion for $\chi$ and $\chi^*$ to get the second equality, and omitted carets on the variables.
Note that we did not take a unitary gauge because the electroweak gauge symmetry is likely to be restored due to thermal corrections from standard model particles before the scalaron decay.\footnote{Even though the universe has not been thermalized before the scalaron decay, homogeneous Higgs condensate \cite{Kunimitsu:2012xx} that is formed during inflation starts to oscillate soon after inflation and decays into quarks and gauge fields, with which a Higgs boson is in thermal equilibrium and acquires thermal mass $\sim \sqrt{\lambda_{\chi}}T \approx 0.1g_*^{-1/4}\sqrt{\lambda_{\chi}} H_{\rm osc} \ll \Gamma(\chi \to {\rm SM\ particles})$ larger than the electroweak scale during the oscillating stage of scalaron. As a result, the electroweak gauge symmetry is restored during preheating. Since the energy density of the relativistic particles is subdominant and decreases faster than $\rho_{\phi}\sim M_p^2H_{\rm osc}^2 \propto a^{-3}$, there is no sizable back-reaction on the dynamics of scalaron.} 
Otherwise, it is convenient to take a unitary gauge with massive gauge bosons.

Based on the above gravitationally induced couplings, the scalaron decay rate is given by \cite{Watanabe:2006ku,Watanabe:2010vy} (also see \cite{Gorbunov:2010bn,Watanabe:2013lwa})
\begin{align}
\Gamma_{\rm tot} = \Gamma(\phi&\to \chi^+\chi^-) + \Gamma(\phi\to\bar\psi\psi) + \Gamma(\phi\to 2A_{\mu}),\nonumber\\
 \Gamma(\phi\to \chi^+\chi^-) &= \frac{N_{\chi}\left[M^2(1+6\xi)+2m_\chi^2\right]^2}{96\pi MM_p^2}\sqrt{1-\frac{4m_\chi^2}{M^2}},\label{eq:decay_scalar}\\
 \Gamma(\phi\to \bar\psi\psi) &=\frac{N_{\psi}Mm_\psi^2}{48\pi M_p^2}\left(1-\frac{4m_\psi^2}{M^2}\right)^{3/2},\label{eq:decay_spinor}\\
 \Gamma(\phi\to 2A_{\mu})&= \frac{N_{A}M^3}{192\pi M_p^2}\left[\frac{\alpha}{\sqrt8 \pi}\sum_{i={\rm heavy} }b_i \right]^2,\label{eq:decay_anomaly}
 \end{align}
where $N_\chi$, $N_\psi$ and $N_A$ are the number of modes for each field. 
In Eq.~(\ref{eq:decay_scalar}), we have included non-minimal gravitational coupling to the Higgs boson $\xi R\chi^*\chi$. 
If $\xi = -1/6$, $\chi$ is conformally coupled to gravity and the induced derivative coupling cancels out; as a result, the leading term in Eq.~(\ref{eq:decay_scalar}) vanishes. 
We shall assume a minimal coupling of the Higgs boson to gravity ($\xi = 0$) to avoid complexity for now. 
As is well known, massless fermions are conformally invariant and the decay rate to a pair of massless fermions vanishes [see Eq.~(\ref{eq:decay_spinor})]. 
These rates are consistent with the Jordan frame analysis~\cite{Starobinsky:1981vz,Vilenkin:1985md, Arbuzova:2011fu}.

Although the scalaron cannot decay into gauge fields classically, it does quantum mechanically via the gauge trace anomaly process with the rate of Eq.~(\ref{eq:decay_anomaly}), where
$\alpha = g^2/(4\pi)$ and $b_i$'s are the lowest coefficients of the beta functions from charged particles \textit{heavier than} the scalaron. 
In the original setup~\cite{Starobinsky:1980te}, the gravitational trace anomaly induces the $R^2$ term whose dimensionless constant $M_p^2/M^2\sim {\cal O}(10^{10})$ is required to match with the observed amplitude of primordial curvature perturbations, which would naively imply the excessive number of degrees of freedom $N_{\rm grav}\sim {\cal O}(10^{10})$  (also see \cite{Bamba:2014jia,Watanabe:2007tf}).
Even if a tiny fraction of $N_{\rm grav}$ is charged under the standard model gauge group, we can expect $|\sum_{i={\rm heavy}}b_i| \sim {\cal O}(10^2) - {\cal O}(10^4)$ at reheating while avoiding strong couplings at inflationary energy scales. Since the gauge coupling constants take values of $\alpha/(\sqrt8 \pi)\sim {\cal O}(10^{-2})-{\cal O}(10^{-1})$ at the energy scale $\sim {\cal O}(10^9 \ {\rm GeV)}$ \cite{Degrassi:2012ry}, the anomaly process could dominate over the scalar channel.
In this case, the shadow of heavy charged particles determines the reheating process~\cite{Watanabe:2010vy}.
Note that this apparently controversial property is similar to the gauge trace anomaly of quantum chromodynamics~\cite{Shifman:1988zk} and to the super-Weyl-K\"ahler anomaly~\cite{Bagger:1999rd}.

If there is no charged particle heavier than the scalaron at reheating, then the anomaly process cannot be expressed by the local effective Lagrangian [the last term of Eq.~(\ref{eq:lag_matt})].
In this case, the anomalous decay rate has to be computed directly from loop diagrams with light intermediate charged particles as in \cite{Watanabe:2010vy}.

Now we can estimate the reheating temperature by using the scalaron decay rate (or equivalently lifetime) as
\begin{align}
T_{\rm rh} &= \frac{\sqrt{\Gamma_{\rm tot}M_p}}{(10\pi^2)^{1/4}}\left(\frac{g_*}{100}\right)^{-\frac14}\nonumber\\
&\simeq M\sqrt{\frac{M}{M_p}} \frac{\sqrt{N_\chi + 2 N_A \left[\frac{\alpha}{\sqrt8 \pi}\sum_i b_i \right]^2}}{ 8\pi (90)^{1/4}}\left(\frac{g_*}{100}\right)^{-\frac14} \nonumber\\
&\sim 10^{-9} M_p,
\end{align}
where $g_*=g_*(T_{\rm rh})$ is the effective number of relativistic species at the time of reheating and the reheating temperature $T_{\rm rh}$ is defined by the moment: $\Gamma_{\rm tot} = 3 H = 3\pi g_*^{1/2}T_{\rm rh}^2/(\sqrt{10}M_p)$. 
To get $T_{\rm rh}= {\cal O}(10^9\ {\rm GeV})$, we have assumed the standard model $g_*=106.75$ as the matter sector and that the scalar decay channel is dominant with $N_\chi =4$ and $\xi =0$ for the Higgs boson.
If the anomaly channel is dominant, the reheating temperature can be as high as $\sim {\cal O}(10^9-10^{12}\ {\rm GeV})$ depending on the number of heavy charged modes.

Since the scalaron oscillation phase evolves as a dust-dominated phase, the number of efolds at the CMB scale and the reheating temperature are associated by \cite{Liddle:2000cg}
\begin{align}
N_* \simeq 54 +\frac13\ln{\left(\frac{T_{\rm rh}}{10^9 \ {\rm GeV}}\right)},
\end{align}
which completes the theoretical predictions from the $R^2$ inflation.

In the following sections, we analyze the dynamics of the scalaron oscillations during the post-inflationary epoch in the Einstein frame and see whether the predictions in this section are modified or not.

\begin{figure*}[htbp]
\begin{center}
\begin{tabular}{c c}
\resizebox{80mm}{!}{\includegraphics{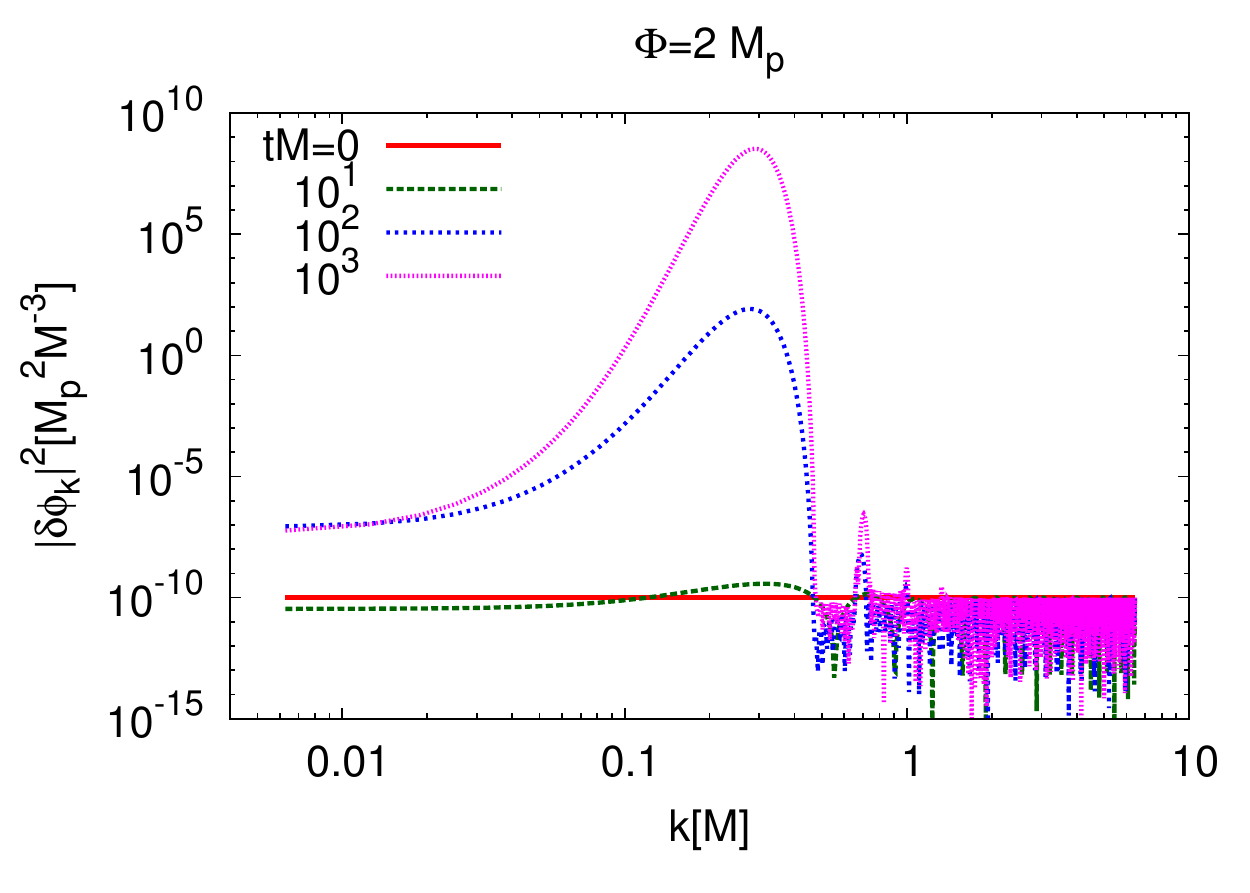}} &
\resizebox{80mm}{!}{\includegraphics{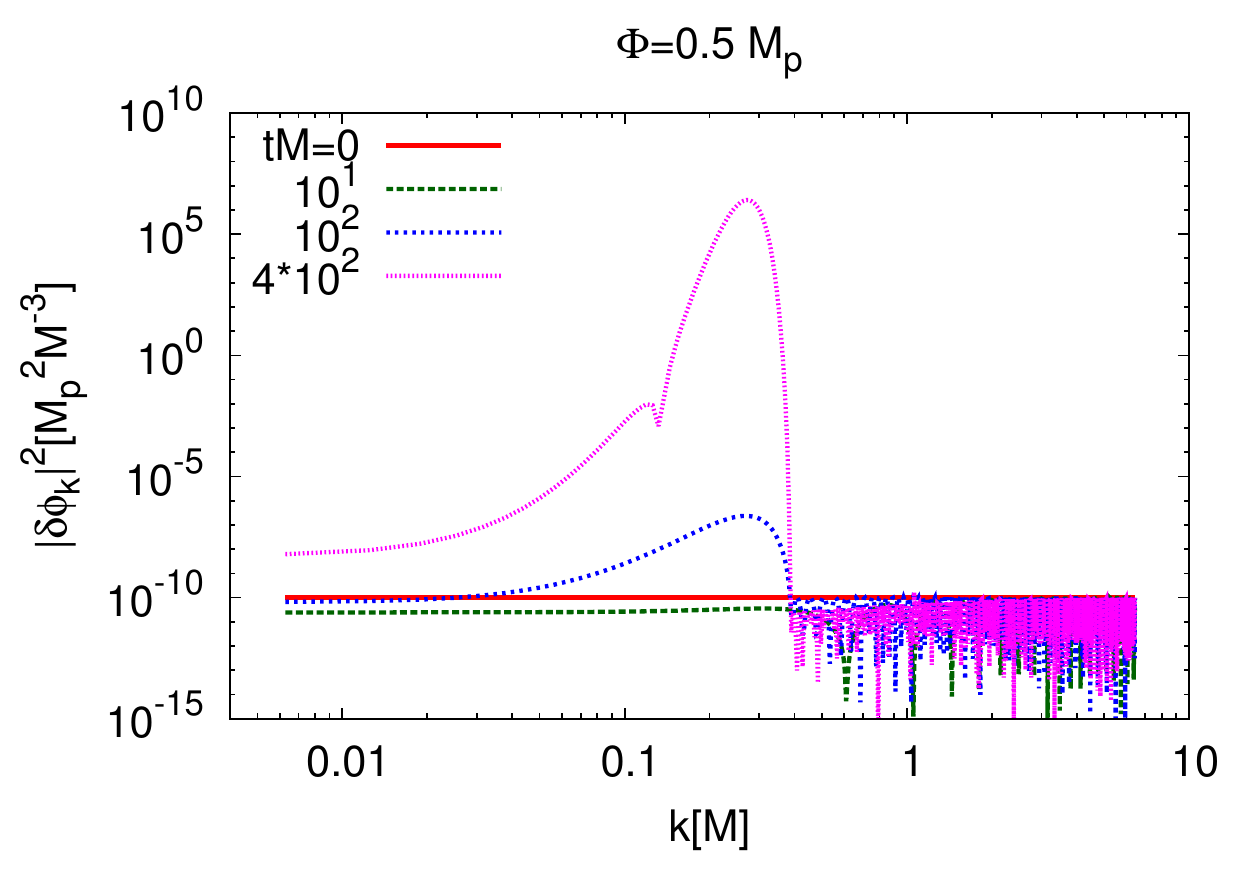}}\\
\resizebox{80mm}{!}{\includegraphics{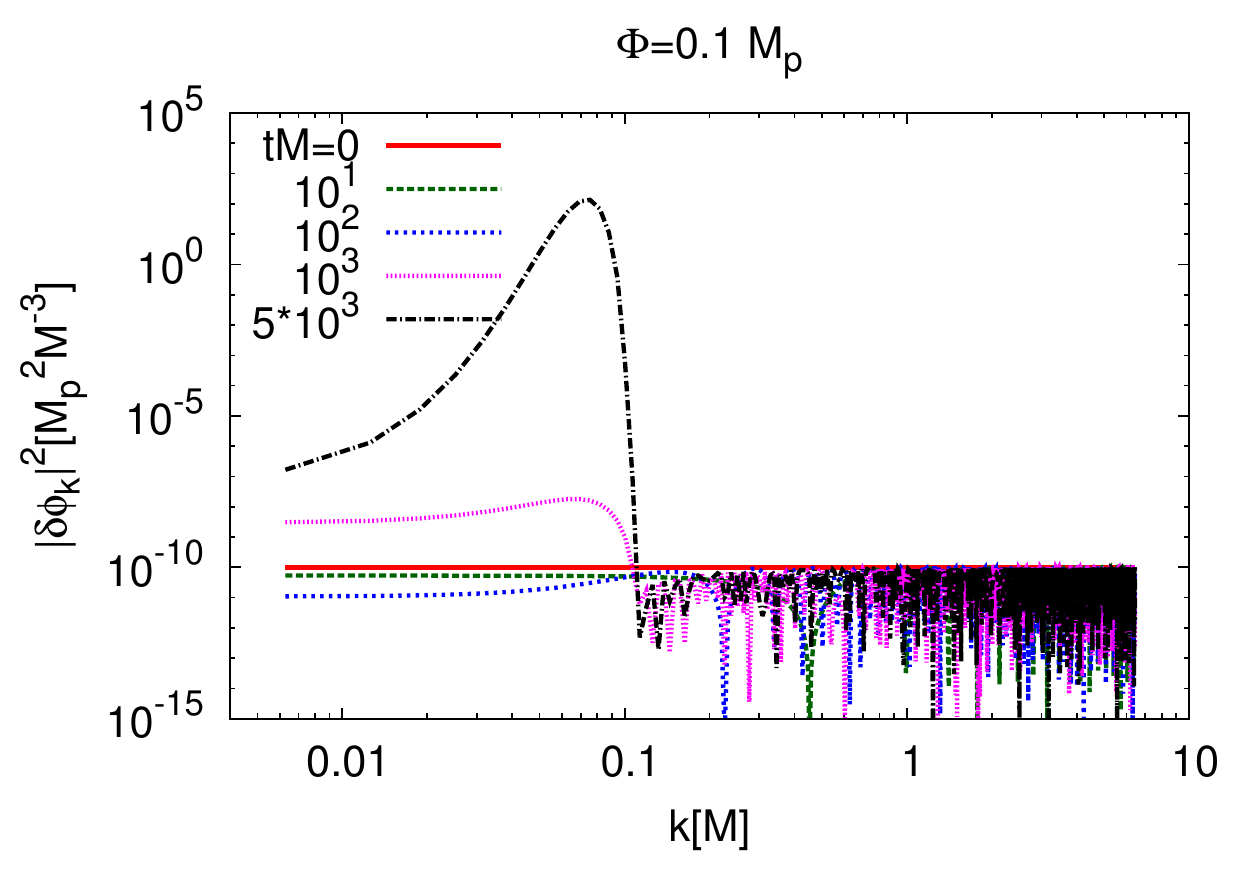}} &
\resizebox{80mm}{!}{\includegraphics{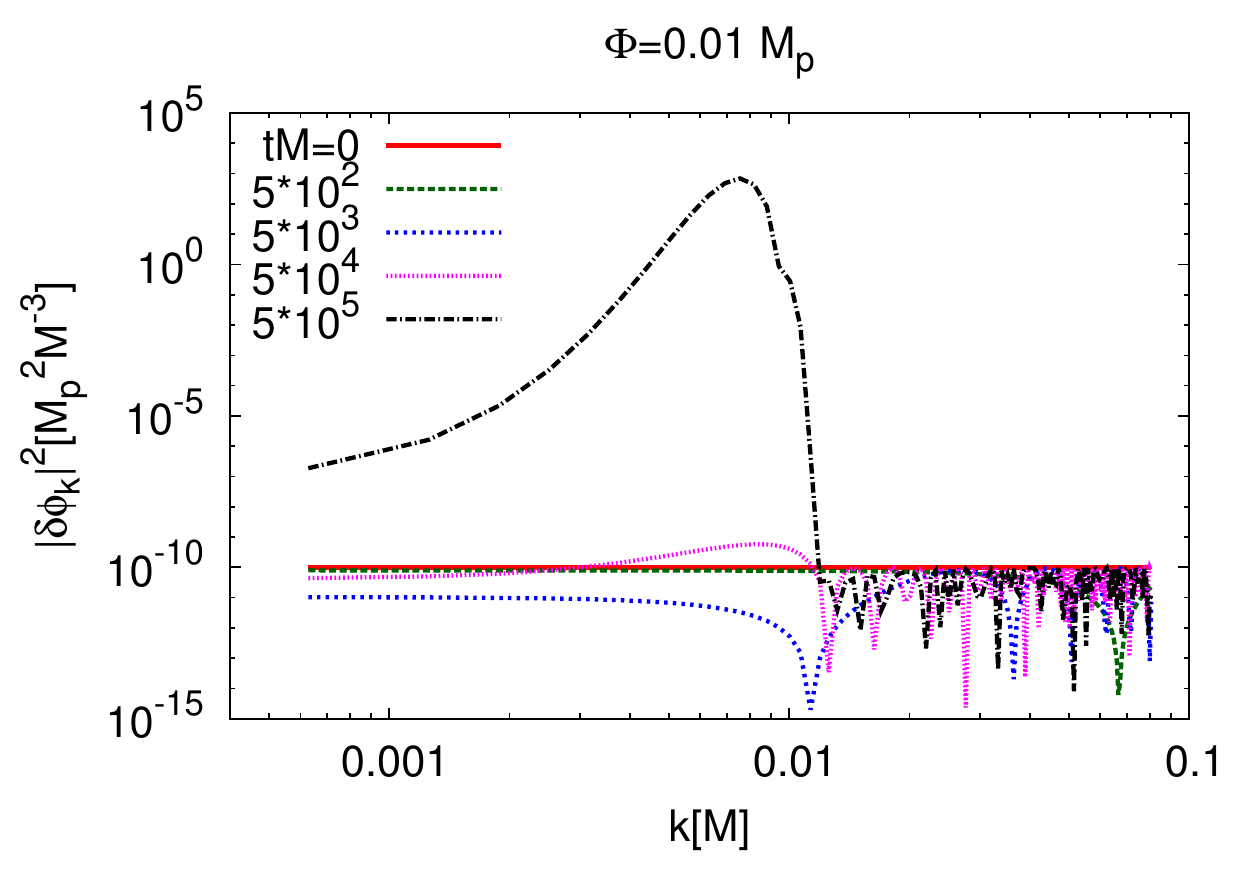}}\\
\end{tabular}
\caption{
Time evolution of the fluctuations from $t=0[1/M]$ for each initial amplitude of the background field 
$\Phi/M_p=2,0.5,0.1,0.01$.
Each line is the snapshot of the spectrum of the fluctuations at several times.
The vertical axis $\delta\phi_k$ is the Fourier mode of the fluctuation $\delta\phi(x)$ and the horizontal 
axis $k$ is the corresponding momentum.
}
\label{FIG_1}
\end{center}
\end{figure*}

\section{Instability ($H=0$)}
\label{Sec_insH0}

In order to form soliton-like objects, oscillons/I-balls, fluctuations of the scalaron field value must grow comparable to its background value.
Here, we study the growth of linear fluctuations in a static Minkowski spacetime ($H=0$) both numerically and analytically. 

In a static spacetime, the equation of motion (EOM) for the scalaron is given by
\begin{equation}\label{eq_insH0_1}
\ddot{\phi}({\bf x},t)-\nabla^2\phi({\bf x},t)+V'(\phi)=0,
\end{equation}
with Eq.~(\ref{eq:pot}).

Dividing $\phi({\bf x},t)$ into the background and fluctuations as $\phi({\bf x},t)=\phi_0(t)+\delta\phi({\bf x},t)$, 
we rewrite the EOM for $\phi_0$ and $\delta\phi$ as
\begin{eqnarray}
\label{eq_insH0_3}
&&\ddot{\phi}_0+V'(\phi_0)=0,\\
\label{eq_insH0_4}
&&\delta\ddot{\phi}_k+k^2\delta\phi_k+V''(\phi_0)\delta\phi_k=0,
\end{eqnarray}
where 
\begin{eqnarray}
\label{eq_insH0_5}
V'(\phi)&=&\sqrt{\frac{3}{2}}M^2M_p\left[1-{\rm e}^{-\sqrt{\frac23}\frac{\phi}{M_p }}\right]{\rm e}^{-\sqrt{\frac23}\frac{\phi}{M_p}},\\[1ex]
\label{eq_insH0_6}
V''(\phi)&=&M^2\left[-1+2{\rm e}^{-\sqrt{\frac23}\frac{\phi}{M_p}}\right]{\rm e}^{-\sqrt{\frac23}\frac{\phi}{M_p}}.
\end{eqnarray}
We have defined the mode function $\delta\phi_k$ by promoting $\delta\phi$ to an operator $\widehat{\delta\phi}$ in the Heisenberg picture as
\begin{align}
\widehat{\delta\phi}({\bf k}) =& \int d^3x~\widehat{\delta\phi}({\bf x}){\rm e}^{-i{\bf k}\cdot{\bf x}}\nonumber\\
=& \delta\phi_k \hat{a}({\bf k})+ \delta\phi_{-k}^*\hat{a}^\dagger(-{\bf k}),
\end{align}
where $\hat{a}({\bf k})$ and $\hat{a}^\dagger({\bf k})$ are annihilation and creation operators with $[\hat{a}({\bf k}), \hat{a}^\dagger({\bf k}')]=(2\pi)^3\delta^3({\bf k}-{\bf k}')$ and zero for other commutators.

\subsection{Numerical result}\label{sub_nusi}

We integrate the coupled Eqs.~(\ref{eq_insH0_3}) and~(\ref{eq_insH0_4}) numerically with the Runge-Kutta $4$-th order.
In Eq.~(\ref{eq_insH0_4}), the frequency of the fluctuations, $\omega_k^2=k^2+V''(\phi_0)$, depends on the background field $\phi_0$, which would amplify fluctuations if modes are in resonance bands. 
We set the initial values as $\Phi\equiv \phi_0(t_0)=2, \ 0.5, \ 0.1, \ 0.01 \ [M_p]$ for the background field, $\dot{\phi}_0(t_0)=0$ for the initial field velocity, and $\delta\phi_k(t_0)=10^{-5}~[M_pM^{-3/2}]$ for every $k$-mode of the field fluctuations.
Since $\phi_0(t_f)\approx 0.94 M_p$ at the end of inflation, these initial conditions will be relevant as time evolves. 

Fig.~\ref{FIG_1} shows the time evolution of the fluctuations 
for different $\Phi$. 
Each panel shows a power spectrum of $\delta\phi_k$ at four or five time slices.
For $\Phi\gtrsim 0.1M_p$, the fluctuations are amplified 
around a particular scale $k\sim 0.1~M$, and for $\Phi\lesssim 0.01M_p$ they are amplified around $k\sim 0.01~M$. 
We find from Fig.~\ref{FIG_1} that the typical time scale of the amplification, if it happens,
is larger than hundred unit times, $\Delta t \gtrsim 100/M$.
Thus, the scalaron self-interactions induce instabilities that grow after teens of oscillations.

 This self-resonance of scalaron fluctuations can be understood analytically by the Floquet solutions of the Mathieu equation~\cite{Abramowitz} as in the standard analysis of parametric resonance~\cite{Kofman:1994rk, Landau}. 

\subsection{Analytical understanding}\label{sub_ana}

Expanding the exponential factor of the potential term in Eq.~(\ref{eq_insH0_3}) as
$V'\simeq M^2\phi_0 + \cdots$, 
we approximate the background solution as a harmonic oscillator with frequency $M$: $\phi_0(t)\simeq\Phi\cos(Mt)$, where 
$\Phi \simeq$ constant.
This background oscillation leads to the time-dependent frequencies of the fluctuations, and eventually to instabilities. 
To clarify self-resonance bands, we expand Eq.~(\ref{eq_insH0_6}) up to quadratic terms as
\begin{equation}
\label{eq_sb_Anau_1}
V''
\simeq M^2\left[1-\sqrt6\frac{\phi}{M_p}
+\frac{7}{3}\left(\frac{\phi}{M_p}\right)^2\right].
\end{equation}
Then Eq.~(\ref{eq_insH0_4}) reads
\begin{equation}
\label{eq_sb_Anau_2}
\delta\ddot{\phi}_k+\omega_k^2\delta\phi_k=0,
\end{equation}
where
\begin{align}
\label{eq_sb_Anau_3}
\omega_k^2=&k^2+M^2\left[1+\frac{7}{6}\left(\frac{\Phi}{M_p}\right)^2\right] \nonumber\\
&-\sqrt6 M^2\frac{\Phi}{M_p}\cos(Mt) 
+\frac{7}{6}M^2\left(\frac{\Phi}{M_p}\right)^2\cos(2Mt) .
\end{align}
For smaller $\Phi$, the time dependence of the frequency~(\ref{eq_sb_Anau_3}) is dominated by $\cos{(Mt)}$, while for the larger $\Phi$, it is also contributed by a higher harmonic, $\cos{(2Mt)}$. 
We parametrize a contribution of the higher harmonic by taking a ratio of the two coefficients:
\begin{equation}
\label{eq_sb_Anau_4}
\alpha
\equiv\frac{7}{6\sqrt{6}}\frac{\Phi}{M_p}.
\end{equation}

When $\alpha \lesssim 1$, the expansion of the potential makes sense.
In this case, $\Phi \lesssim 2 M_p$ and the lower harmonic dominates the frequency:
\begin{equation}
\label{eq_sb_Anau_10}
\omega_k\simeq 
k^2+M^2\left[1+\frac{7}{6}\left(\frac{\Phi}{M_p}\right)^2\right]
-\sqrt{6}M^2\frac{\Phi}{M_p}\cos(Mt).
\end{equation}
Redefining the time variable as $Mt\equiv 2\hat{T}$, we can transform the EOM for $\delta\phi_k$ [Eq.~(\ref{eq_insH0_4})] into the Mathieu equation:
\begin{equation}
\label{eq_sb_Anau_11}
\delta\phi_k''+\left[A_{1k}-2q_1\cos(2\hat{T})\right]\delta\phi_k=0,
\end{equation}
where a prime denotes a derivative with respect to $\hat{T}$. Parameters $q_1$ and $A_{1k}$ are defined as
\begin{eqnarray}
\label{eq_sb_Anau_12}
q_1&\equiv& 2\sqrt{6}\frac{\Phi}{M_p},\\
\label{eq_sb_Anau_13}
A_{1k}&\equiv& 4 + 4\left(\frac{k}{M}\right)^2+\frac{7}{36}q_1^2.
\end{eqnarray}
The Mathieu equation~(\ref{eq_sb_Anau_11}) has a growing mode solution that leads to instabilities.
The instabilities are classified to two cases, whether $q<1$ or $q>1$.
When $q<1$, a growth of the fluctuations can be induced by a narrow resonance~\cite{Shtanov:1994ce} whose instability 
mode is characterized by the parameter $A_k=n^2$, where $n$ is a natural number.
When $q>1$, a growth of the fluctuations can be induced by a broad resonance~\cite{Kofman:1994rk}; namely, instabilities occur 
at the break down of the adiabatic condition: $|d\omega/dt|/\omega^2>1$.

\begin{figure}[tbp]
\begin{center}
\begin{tabular}{c}
\resizebox{90mm}{!}{\includegraphics{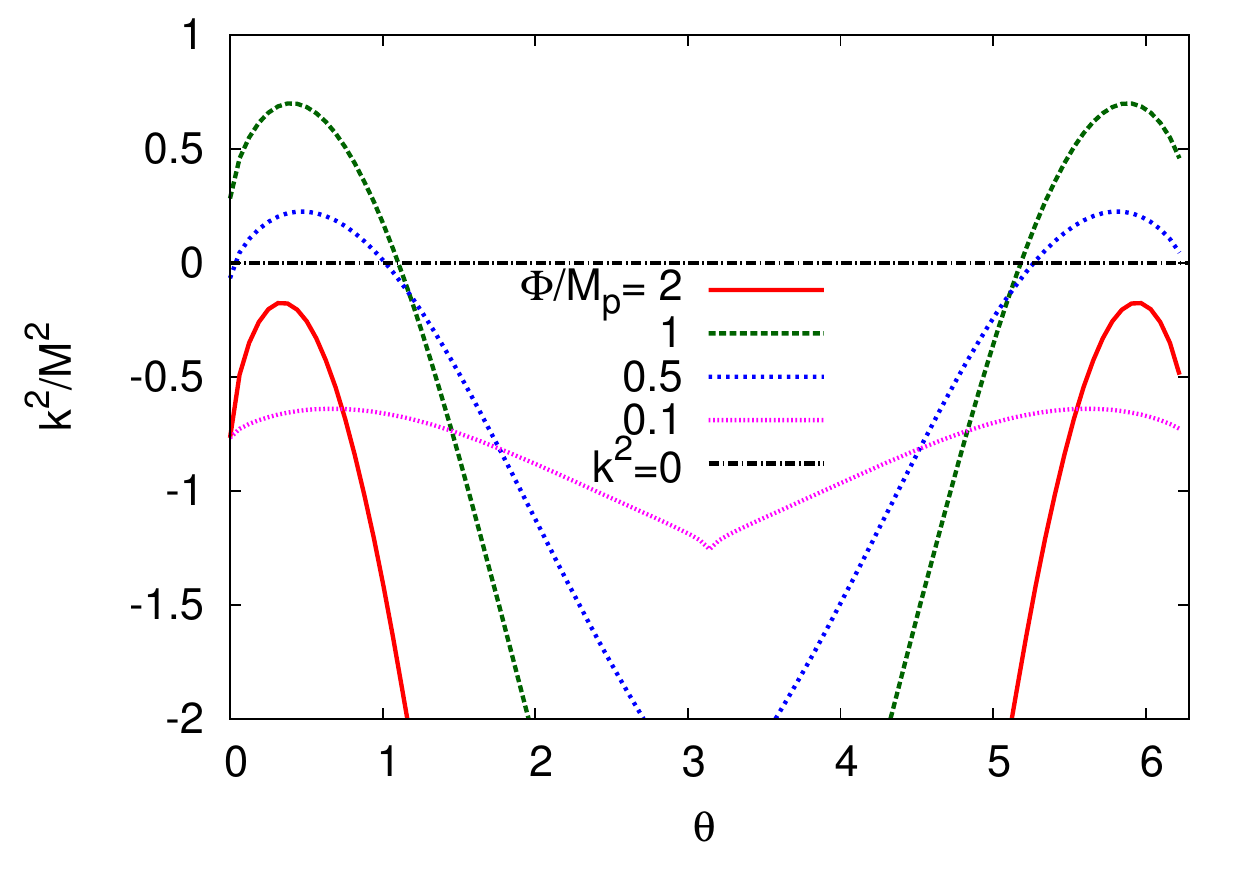}} 
\end{tabular}
\caption{
Instability modes of the broad resonance for $\alpha<1$ and initial amplitudes 
$\Phi/M_p=2,1,0.5,0.1$. 
The vertical axis is the right-hand side of the resonance condition~(\ref{eq_sb_Anau_14}).
The horizontal axis is the time variable which is defined as $\theta\equiv Mt~({\rm modulo~2\pi})$.
The enhancement of the fluctuations occurs in $k^2/M^2>0$ region.
}
\label{FIG_3}
\end{center}
\end{figure}

For $0.2 M_p \lesssim \Phi \lesssim 2 M_p$ such that $q_1 \gtrsim 1$, enhancement of the fluctuations would be induced by a broad resonance. 
Substituting the frequency~(\ref{eq_sb_Anau_10}) into the non-adiabatic condition $|d\omega/dt|/\omega^2>1$, we get
\begin{align}
\label{eq_sb_Anau_14}
\left(\frac{k}{M}\right)^2<& -1-\frac{7}{6}\left(\frac{\Phi}{M_p}\right)^2 
+\sqrt{6}\frac{\Phi}{M_p}\cos(Mt) \nonumber\\
&+\left(\frac{3}{2}\right)^{\frac13}\left(\frac{\Phi}{M_p}\right)^{\frac23}\left|\sin(Mt)\right|^{\frac23},
\end{align}
which yields a broad resonance band.
Fig.~\ref{FIG_3} shows the right-hand side of Eq.~(\ref{eq_sb_Anau_14}). 
An exponential instability takes place if the adiabatic condition is violated, i.e. if the lines in Fig.~\ref{FIG_3} lie above zero for some time.
We find the instability can occur for $\Phi = 0.5 M_p$ within $0 \le k/M < 0.47$, which explains the broad resonance band of the right top panel in Fig.~\ref{FIG_1}.
For $\Phi=2M_p$, even though $q_1 = 4\sqrt6 > 1$, the broad resonance condition [Eq.~(\ref{eq_sb_Anau_14}) and Fig.~\ref{FIG_3}] indicates that there is no instability band due to the lower harmonic frequency. In this case, however, the higher harmonic becomes important, which we will explain later.

For $\Phi<0.2M_p$, parameter $q_1$ is smaller than 1, and a growth of the fluctuations can be induced instead by narrow resonance. 
The narrow resonance condition for the Mathieu equation~(\ref{eq_sb_Anau_11}) is given by \cite{Abramowitz, Landau}
\begin{align}\label{eq:narrow}
- \frac{q^n}{n^{n-1}} \lesssim A_{k}-n^2 \lesssim \frac{q^n}{n^{n-1}},
\end{align}
where $n$ is a natural number. 
The width of parametric resonance bands rapidly decreases as $q^n$ for large $n$, and the rate of exponential growth does as well.  
Thus, lower resonance bands are more important.

Since there is no real solution for $n=1$ and $A_k = A_{1k}$, the second instability band is the most important.
In this case, the condition~(\ref{eq:narrow}) can be given more precisely by (see Sec.~20.2.25 of \cite{Abramowitz})
\begin{align}
-\frac{q^2}{12} < A_k - 4 < \frac{5q^2}{12},
\end{align}
and thus for $q=q_1$ and $A_k = A_{1k}$
\begin{align}\label{eq:second_narrow_minkowski}
0 \le \frac{k}{M} < \frac{q_1}{3\sqrt2}.
\end{align}
The narrow resonance condition tells that, for $\Phi = 0.1 M_p$ and $0.01 M_p$, the instabilities occur within $0 \le k/M < 1/(5\sqrt3)\approx 0.115$ and $0 \le k/M < 1/(50\sqrt3) \approx 0.0115$, respectively. This explains the growth of the fluctuations in the bottom panels in Fig.~\ref{FIG_1}.

For $\alpha \gtrsim 1$ (namely $\Phi \gtrsim 2 M_p$), the perturbative truncation of the potential~(\ref{eq_sb_Anau_1}) fails.
Still, it is instructive to show how the instabilities in Fig.~\ref{FIG_1}  are induced in this case, by estimating a contribution from the higher harmonic $\cos{(2Mt)}$.
The frequency of the fluctuations is approximated as 
\begin{equation}
\label{eq_sb_Anau_5}
\omega_k\simeq 
k^2+M^2\left[1+\frac{7}{6}\left(\frac{\Phi}{M_p}\right)^2 \right]
+\frac{7}{6}M^2\left(\frac{\Phi}{M_p}\right)^2\cos(2Mt).
\end{equation}
For this frequency, shifting the time as $2Mt\rightarrow 2Mt+\pi$ and redefining the time variable  as $2Mt\equiv 2\hat{T}$, 
we can transform the EOM for $\delta\phi_k$ as
\begin{equation}
\label{eq_sb_Anau_6}
\delta\phi''_k+\left[A_{2k}-2q_2\cos(2\hat{T})\right]\delta\phi_k=0,
\end{equation}
where a prime denotes a derivative with respect to $\hat{T}$. Parameters $q_2$ and $A_{2k}$ are defined as
\begin{eqnarray}
\label{eq_sb_Anau_7}
q_2&\equiv&\frac{7}{12}\left(\frac{\Phi}{M_p}\right)^2,\\
\label{eq_sb_Anau_8}
A_{2k}&\equiv&1+\left(\frac{k}{M}\right)^2+2q_2.
\end{eqnarray}
For $\Phi \ge 2 M_p$ (i.e. $\alpha \ge 7/(3\sqrt{6})\approx 0.95$), the parameter $q$ is larger than 1 as
$q_2 \ge 7/3 \approx 2.3$. 
Thus, instabilities can be induced by broad resonance due to the second harmonic $\cos{(2Mt)}$ [and a linear combination of higher harmonics $\cos{(nMt)}$].

\begin{figure}[tbp]
\begin{center}
\begin{tabular}{c}
\resizebox{90mm}{!}{\includegraphics{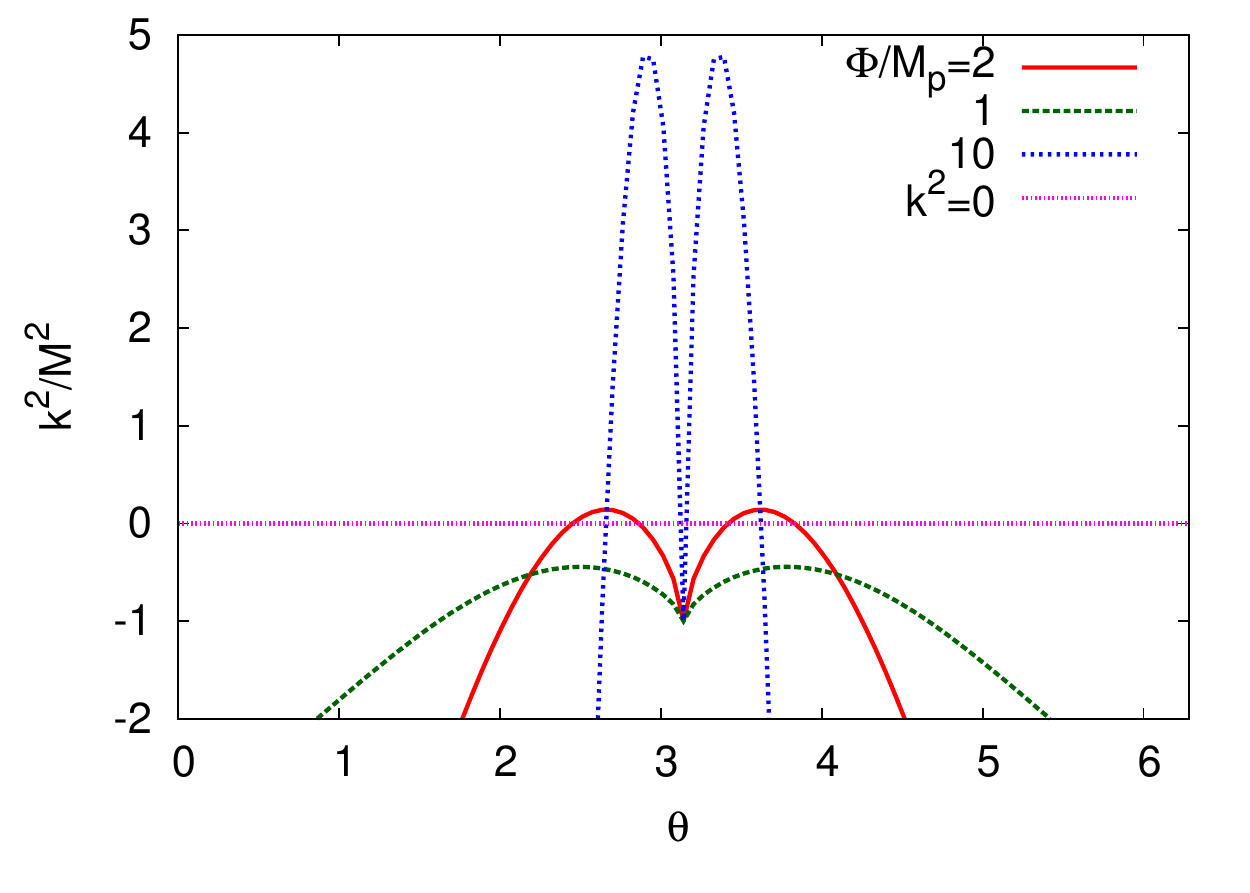}} 
\end{tabular}
\caption{
Instability modes of the broad resonance for $\alpha > 1$ and initial amplitudes 
$\Phi/M_p=10,2,1$. 
The vertical axis is the right-hand side of the resonance condition~(\ref{eq_sb_Anau_9}).
The horizontal axis is the time variable which is defined as $\theta\equiv 2Mt~({\rm modulo~2\pi})$.
The enhancement of the fluctuations occurs in $k^2/M^2>0$ region.
}
\label{FIG_2}
\end{center}
\end{figure}

\begin{figure}[tbp]
\begin{center}
\begin{tabular}{c}
\resizebox{90mm}{!}{\includegraphics{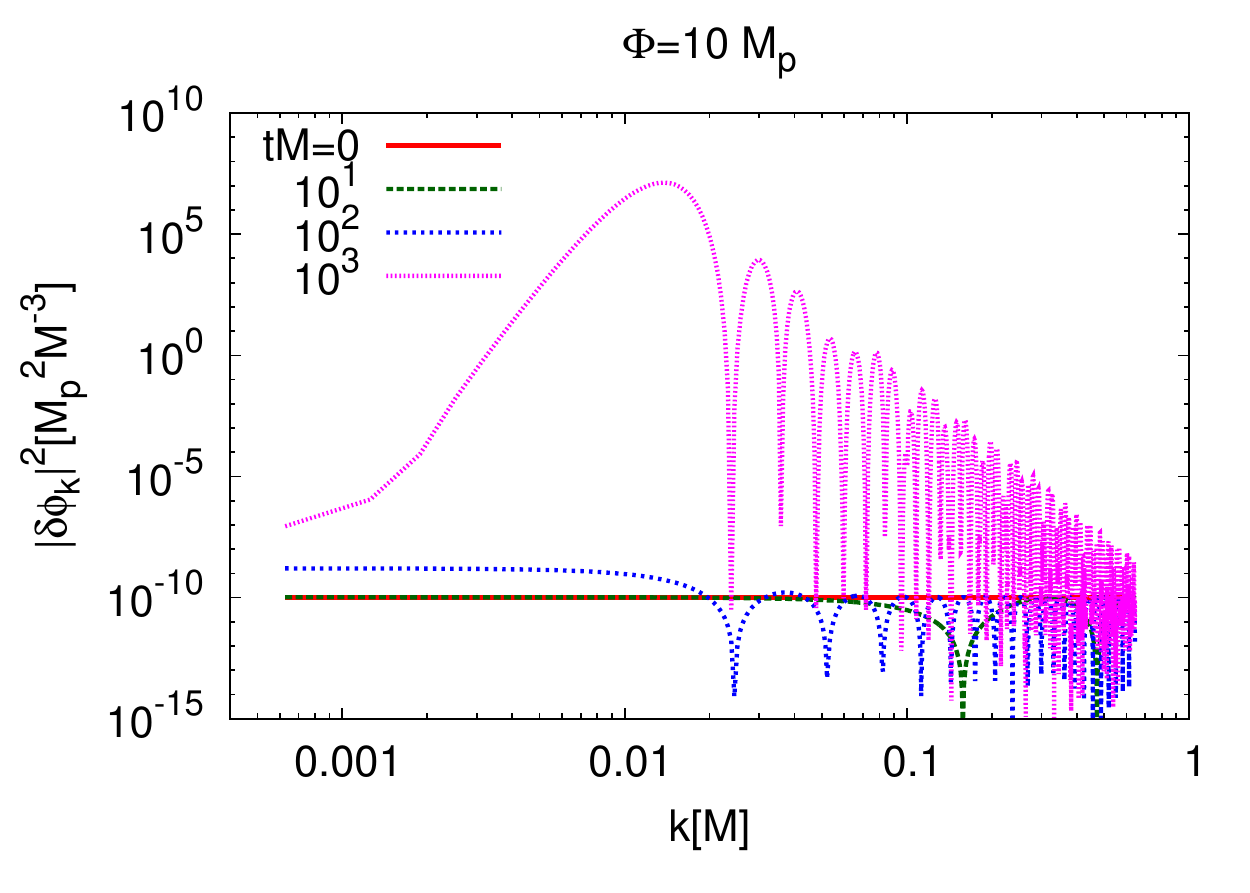}} 
\end{tabular}
\caption{
Time evolution of the fluctuations from $t=0[1/M]$ to $t=10^3[1/M]$ for 
$\Phi/M_p=10$.
Each line is the snapshot of the spectrum of the fluctuations at several times.
The vertical axis $\delta\phi_k$ is the Fourier mode of the fluctuation $\delta\phi(x)$ and the horizontal 
axis $k$ is the corresponding momentum.
}
\label{FIG_7}
\end{center}
\end{figure}

Substituting the frequency~(\ref{eq_sb_Anau_5}) into the non-adiabatic condition $|d\omega/dt|/\omega^2>1$, we get 
the following inequality
\begin{align}
\label{eq_sb_Anau_9}
\left(\frac{k}{M}\right)^2< &
-1-\frac{7}{6}\left(\frac{\Phi}{M_p}\right)^2[1+\cos(2Mt)] \nonumber\\
&+\left(\frac{7}{6}\right)^{\frac23}\left(\frac{\Phi}{M_p}\right)^{\frac43}\left|\sin(2Mt)\right|^{\frac23}.
\end{align}
The $k$-modes, that satisfy the above condition~(\ref{eq_sb_Anau_9}), exponentially grow with respect to time. 
Fig.~\ref{FIG_2} shows the right-hand side of the resonance condition~(\ref{eq_sb_Anau_9}). Instabilities occur  if the lines lie above zero for some time.
For $\Phi = 2 M_p$ and $10 M_p$, broad resonances are induced within $0 \le k/M < 0.37$ and $0 \le k/M < 2.2$, respectively.
These $k$-modes explain instabilities in the upper left-hand of Fig.~\ref{FIG_1} and Fig.~\ref{FIG_7}, respectively.

In this section, we have investigated the exponential growth of the scalaron fluctuations with the potential~(\ref{eq:pot}) in 
a static Minkowski spacetime.
We have found that narrow and broad parametric resonances are induced by the background oscillations of the scalaron with self-interactions~(\ref{eq_sb_Anau_3}).
For small amplitudes of the background oscillations $\Phi < 0.2 M_p$, the second band of narrow resonance amplifies the scalaron fluctuations.
On the other hand, for large amplitudes $\Phi \gtrsim 0.2 M_p$, the effects of broad resonance amplify the scalaron fluctuations either by a harmonic $\cos{(Mt)}$ if $\Phi < 2 M_p$ or by a linear combination of higher harmonics $\sum_{n \ge 2}\cos{(n Mt)}$ if $\Phi \gtrsim 2 M_p$.
Therefore, in a Minkowski spacetime, the oscillating scalaron is highly likely to fragment into quasi-stable objects during the post-inflationary epoch.
To confirm the formation of oscillons/I-balls, we have to follow the non-linear dynamics of the field fluctuations by using lattice simulations. 
However, before executing lattice simulations, we must take into account the effects of a more realistic spacetime.

\section{Instability ($H\neq 0$)}
\label{Sec_insHn0}
So far, we have ignored the expansion of the universe in the analysis of parametric resonances. 
Since the expansion stretches and diminishes the fluctuations, preheating may be less efficient in an expanding Friedmann background.

\subsection{Rigid Friedmann background}\label{subsec:friedmann}
At first, for simplicity, we ignore back-reaction from metric perturbations on $\delta\phi_k$. 
In a flat Friedmann-Lema\^{i}tre-Robertson-Walker (FLRW) spacetime, the EOM for $\phi$ is given by
\begin{align}
\label{eq_insHn0_1}
\ddot{\phi}+3H\dot{\phi}-\frac{\nabla^2}{a^2}\phi+V'(\phi)=0,\\
H^2=\frac{1}{3M_p^2}\left[\frac{\dot\phi^2}{2}+V(\phi)\right],\label{eq_insHn0_1_2}
\end{align}
where $a$ is the FLRW scale factor.
Dividing again $\phi$ into the background and fluctuations as $\phi(x,t)=\phi_0(t)+\delta\phi(x,t)$, we rewrite the EOM for 
$\phi_0$ and $\delta\phi$ as 
\begin{eqnarray}
\label{eq_insHn0_2}
&&\ddot{\phi}_0+3H\dot{\phi}_0+V'(\phi_0)=0,\\
\label{eq_insHn0_3}
&&\delta\ddot{\phi}_k+3H\delta\dot{\phi}_k+\frac{k^2}{a^2}\delta\phi_k+V''(\phi_0)\delta\phi_k=0,
\end{eqnarray}
where we have approximately determined the Hubble parameter by the background field as
$H^2\simeq[\dot{\phi}_0^2/2+V(\phi_0)]/(3M_p^2)$.

The onset of preheating can be chosen at the end of inflation. The field value at this time, $t_f$, is given by violation of the slow roll $\epsilon(t_f) =1$, which yields $\phi_0(t_f)=0.94 M_p$. 

In the previous section, we have shown that in the Minkowski space-time ($H=0$) the fluctuations are 
enhanced by the background oscillations with $\Phi \approx M_p$, and it takes $\Delta t \gtrsim 100 \, [1/M]$ to grow significantly.
In the expanding universe, however, since the amplitude of the oscillating background field damps as 
$\phi_0\propto a^{-3/2}\propto t^{-1}$, $\phi_0$ becomes too small to induce the growth of the fluctuations 
after a hundred unit times or a few tens of oscillations: $\phi_0(t_f+\Delta t)\approx 0.94M_p/100 \sim{\cal O}(10^{-2})M_p$. 
Therefore, instabilities due to parametric self-resonance would not grow after $R^2$-inflation.

\begin{figure}[tbp]
\begin{center}
\begin{tabular}{c}
\resizebox{90mm}{!}{\includegraphics{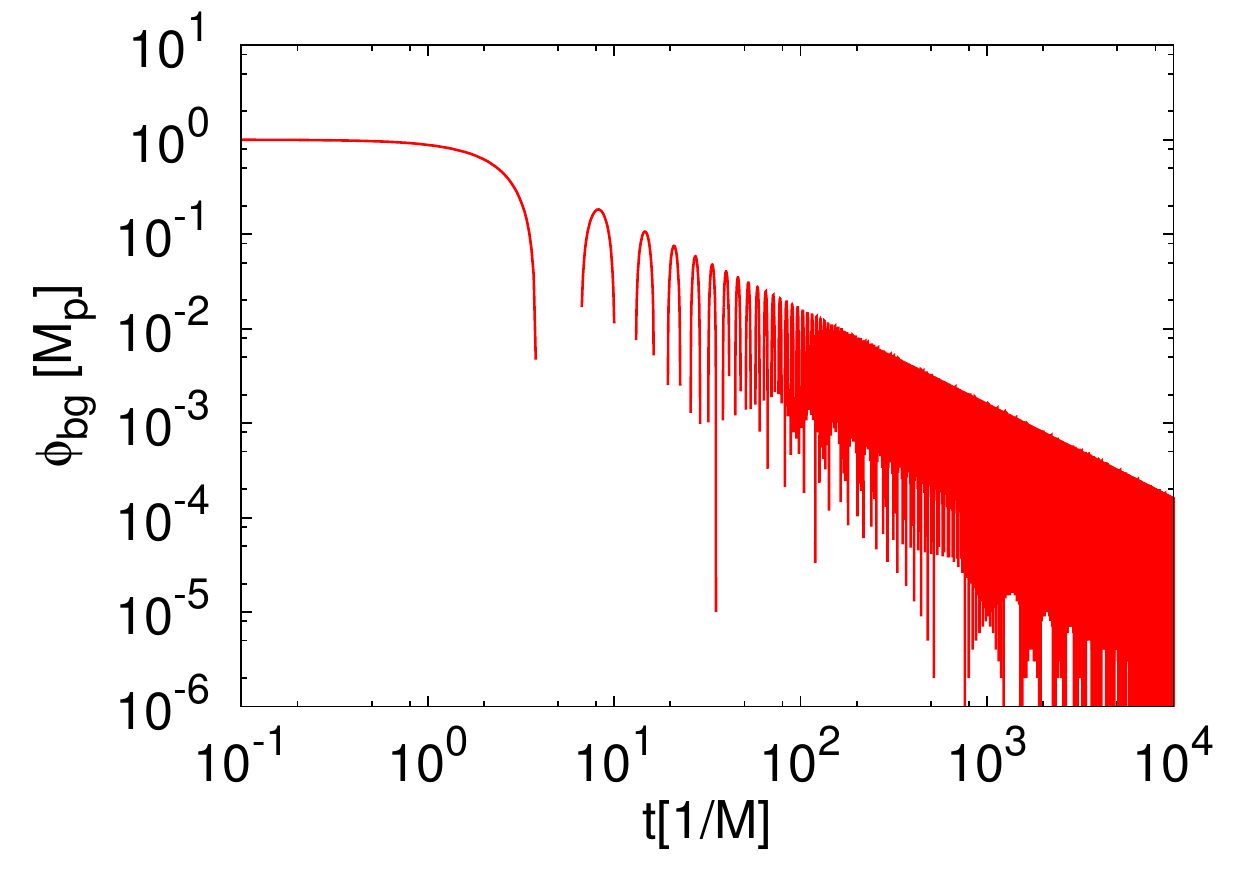}} 
\end{tabular}
\caption{
Time evolution of the background from $t=10^{-1}[1/M]$ to $t=10^4[1/M]$.
The vertical axis $\phi_{\rm bg}=\phi_0$ is the amplitude of the background field,
and the horizontal axis $t$ is cosmic time in units of $M^{-1}$. 
}
\label{FIG_4}
\end{center}
\end{figure}

Fig.~\ref{FIG_4} shows the time evolution of the background field value. We have numerically integrated Eqs.~(\ref{eq_insHn0_1}) and (\ref{eq_insHn0_1_2}) with the Runge-Kutta 4-th order method. We have set the initial field value as $\phi_0(t_{\rm ini})=1M_p$ for definiteness. 
The envelope of the field value $\phi_0$ is indeed decreasing as $t^{-1}$.
We find that inflation ends at $t \approx 4 \ [1/M]$ and $\phi_0 \approx 0.9 M_p$, and that $\phi_0 \sim {\cal O}(10^{-2})M_p$ after a hundred unit times $t \sim 100 \ [1/M]$ as was estimated. 

\begin{figure}[tbp]
\begin{center}
\begin{tabular}{c}
\resizebox{90mm}{!}{\includegraphics{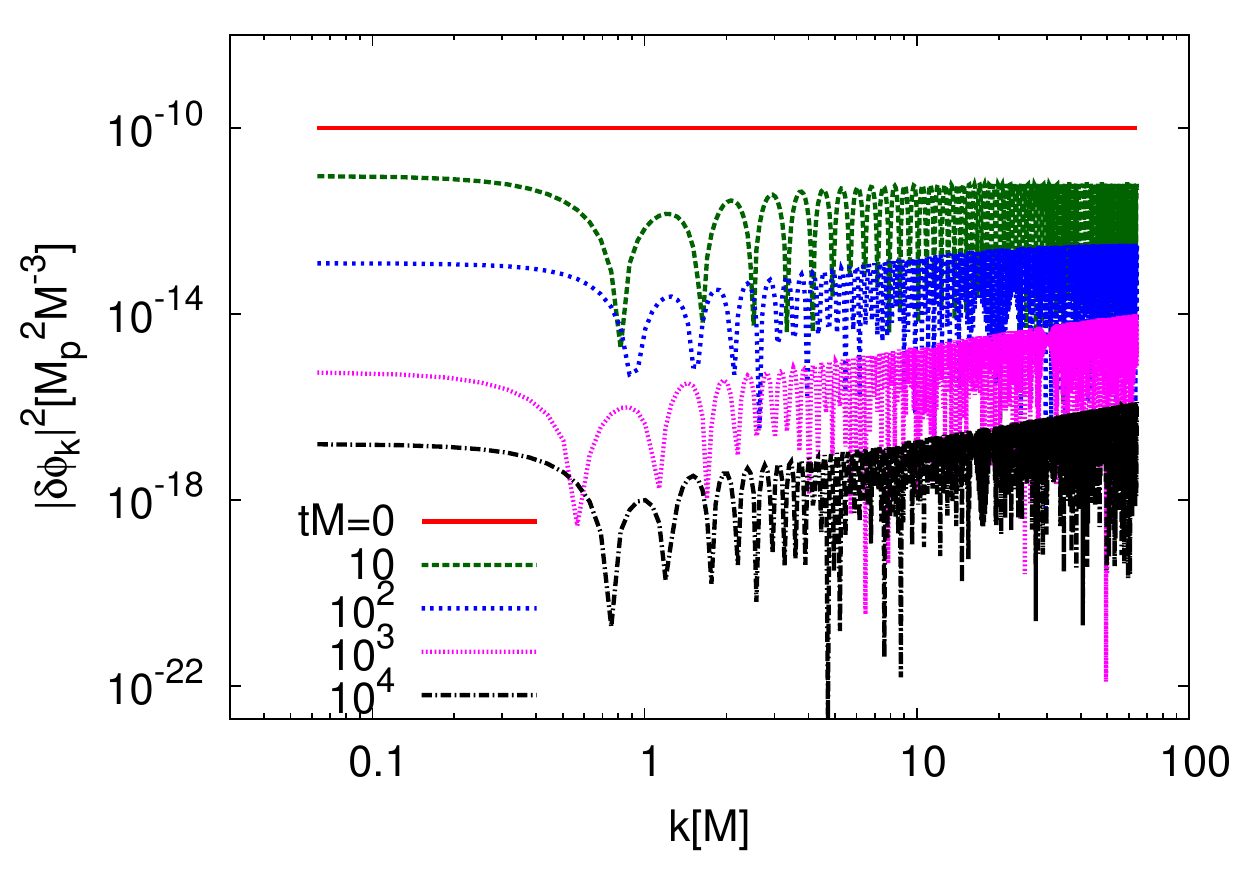}} 
\end{tabular}
\caption{
Time evolution of the fluctuations from $t=0[1/M]$ to $t=10^{4}[1/M]$.
Each line is the snapshot of the spectrum of the fluctuations at
$t=0,10,10^2,10^3,10^4[1/M]$.
The vertical axis $\delta\phi_k$ is the Fourier mode of the fluctuation $\delta\phi(x)$, and the horizontal 
axis $k$ is the corresponding momentum.
}
\label{FIG_5}
\end{center}
\end{figure}

Fig.~\ref{FIG_5} shows that the time evolution of the field fluctuations $\delta\phi_k$ for comoving wave numbers $k$ from $2\pi\times 10^{-2}\ [M]$ to $2\pi\times 10 \ [M]$ which cover self-resonance modes studied in Fig.~\ref{FIG_1}. 
We have set the initial values of the fluctuations as $\delta\phi_k = 10^{-5} \ [M_pM^{-3/2}]$ for every $k$-mode.
We find that fluctuations are indeed decreased by the expansion of the universe. This means that the parametric self-resonances cannot overcome the effect of the expansion, and thus instabilities do not occur during the reheating epoch in the absence of the back-reaction from the metric of space-time.

\subsection{Metric preheating}

We will now take the metric perturbations into account. 
Under the coordinate (gauge) transformation $t\to t+\delta t$, the field and the scale factor are transformed as $\phi(x,t+\delta t) \simeq \phi_0(t)+\delta\phi(x,t)+\dot\phi_0(t) \delta t$ and $a(t+\delta t) \simeq a(t) + \dot{a}(t)\delta t = a(t)+a(t)H \delta t$, respectively.
With an almost flat FLRW spatial metric $h_{ij}(x,t+\delta t)dx^idx^j\simeq a^2(t){\rm e}^{2\zeta}dx_i^2$, we take $H\delta t = \zeta$ and $\phi(x,t+\delta t)\simeq \phi_0(t)+\delta\phi(x,t)+\dot\phi_0(t)\zeta/H$. Then, we cannot separate fluctuations $\delta\phi$ and $\zeta$ since their linear combination is physical.
In this case, the gauge invariant combination of the perturbations is known as the Mukhanov-Sasaki variable \cite{Mukhanov:1985rz}:
\begin{align}
v_k = a\left(\delta\phi_k - \frac{\dot\phi_0}{H}\zeta_k\right),
\end{align}
where $\zeta_k$ is the perturbation of the three curvature given in Sec.~\ref{Sec_model}.
During the post-inflationary oscillation phase, $\zeta_k$ has a singular behavior, which can be avoided  by instead using the EOM for $v_k$ (Mukhanov-Sasaki equation \cite{Mukhanov:1985rz}) \cite{Nambu:1989kh,Kodama:1996jh,Nambu:1996gf,Finelli:1998bu,Jedamzik:2010dq, Easther:2010mr}. In a flat gauge (i.e., on the zero curvature time slice, $\zeta_k=0$), the Mukhanov-Sasaki equation reads
\begin{eqnarray}
\label{eq_insHn0_5}
\delta\ddot{\phi}_k+3H\delta\dot{\phi}_k+\left[\frac{k^2}{a^2}+V''(\phi_0)+\Delta F\right]\delta\phi_k=0,
\end{eqnarray}
where the back-reaction from the metric perturbations is given by
\begin{equation}
\label{eq_insHn0_6}
\Delta F\equiv
\frac{\dot{2\phi_0}}{M_p^2H}V'(\phi_0)
+\frac{\dot{\phi}_0^2}{M_p^4H^2}V(\phi_0).
\end{equation}
The background equations are given by Eqs.~(\ref{eq_insHn0_1_2}) and (\ref{eq_insHn0_2}).

\begin{figure}[tbp]
\begin{center}
\begin{tabular}{c}
\resizebox{90mm}{!}{\includegraphics{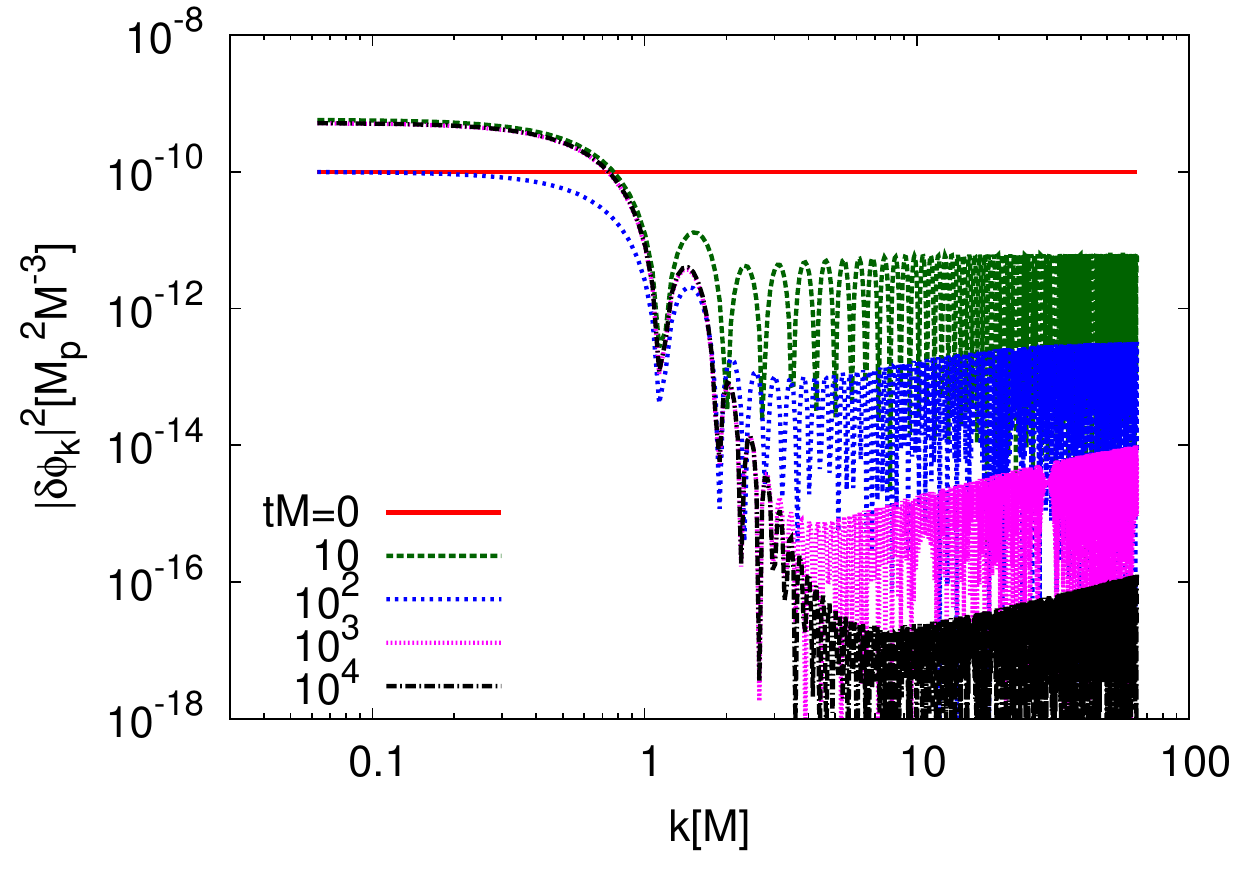}} 
\end{tabular}
\caption{
Time evolution of the fluctuations from $t=0[1/M]$ to $t=10^{4}[1/M]$ with the Mukhanov-Sasaki equation.
Each line is the snapshot of the spectrum of the fluctuations at
$t=0,10,10^2,10^3,10^4[1/M]$.
The vertical axis $\delta\phi_k$ is the Fourier mode of the fluctuation $\delta\phi(x)$ and the horizontal 
axis $k$ is the corresponding momentum.
}
\label{FIG_6}
\end{center}
\end{figure}

We have numerically integrated the coupled Eq.~(\ref{eq_insHn0_1_2}), (\ref{eq_insHn0_2}) and (\ref{eq_insHn0_5}) with the same numerical method as in the previous calculations. 
Initial conditions for the background and fluctuations are set to the same values as in the subsection~\ref{subsec:friedmann}. 

Fig.~\ref{FIG_6} shows the evolution of the fluctuations from $t=0 \ [1/M]$ to $10^4 \ [1/M]$ for comoving wave numbers between 
$k=2\pi\times10^{-2}[M]$ and $2\pi\times10[M]$. 
We can see that the fluctuations are damped for the larger $k$-modes due to the Hubble expansion as in Fig.~\ref{FIG_5}. 
For the smaller $k$-modes, on the other hand, we find growth of the fluctuations. 
The instability occurs near the horizon scale $k\sim 1[M]$, and its growth is balanced with the Hubble expansion. 
This result agrees with the previous works~\cite{Nambu:1996gf,Finelli:1998bu,Jedamzik:2010dq, Easther:2010mr} in which the evolution of the sub-horizon scale Mukhanov-Sasaki variable is calculated during the oscillation epoch with a quadratic potential. 

\begin{figure}[tbp]
\begin{center}
\begin{tabular}{c}
\resizebox{90mm}{!}{\includegraphics{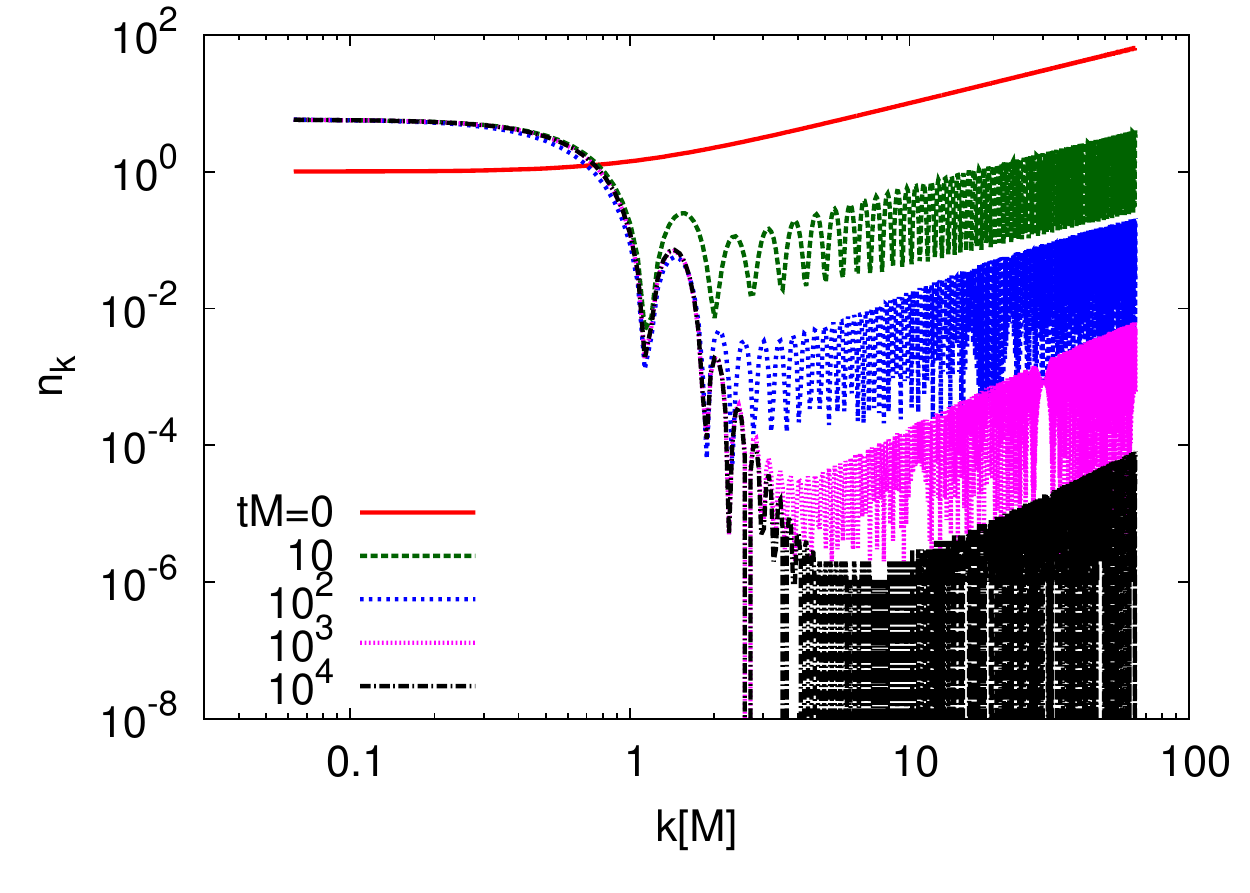}} 
\end{tabular}
\caption{
Time evolution of $n_k$ from $t=0[1/M]$ to $t=10^{4}[1/M]$.
Each line is the snapshot of the spectrum of $n_k$ at
$t=0,10,10^2,10^3,10^4[1/M]$.
The horizontal 
axis $k$ is the comoving momentum corresponding to each $n_k$.
}
\label{FIG_8}
\end{center}
\end{figure}

In Fig.~\ref{FIG_6}, one may notice that there is a decrease of the fluctuations on large scales at $t = 100 \ [1/M]$. 
This damping is caused simply by phase dependence of the fluctuations: $\delta\phi_k \approx f_k(t)\cos(\omega_k t)$ with $\omega_k=\sqrt{k^2+M^2}$, where the amplitude $f_k(t)$ is constant in time. (This phase dependence can be found also in Figs.~\ref{FIG_1},~\ref{FIG_7} and~\ref{FIG_5}.) 
To see this clearly,  we define a phase independent variable as~\cite{Kofman:1994rk}
\begin{equation}
n_k = \frac{\omega_k}{2}\left(\frac{|\delta\dot{\phi}_k|^2}{\omega_k^2}+|\delta\phi_k|^2\right), 
\end{equation}
where we have ignored a contribution from the zero point energy.
This quantity is the phase-independent occupation number of the created particles $\delta\phi_k$ whose time evolution is shown in Fig.~\ref{FIG_8}.
The number of the created particles jumps to 5 at the first half-oscillation and stays constant afterward for $k \lesssim M$.

\subsection{Analytical understanding}
The balance between parametric resonance and  Hubble damping for scales $k\lesssim M$ can be understood as follows.
 The background Eq.~(\ref{eq_insHn0_2}) can be solved approximately as
 \begin{align}
 \phi_0(t) \simeq \phi_0(t_{\rm ini})\left(\frac{a_{\rm ini}}{a}\right)^{\frac32}\sin{(Mt)},
 \end{align}
 which is consistent with Fig.~\ref{FIG_4}.
By plugging the approximate solution into the perturbed Eq.~(\ref{eq_insHn0_5}), we find the frequency:
\begin{align}\label{eq:omega_metric}
\omega_k^2 \simeq \frac{k^2}{a^2} +M^2\left(1-\sqrt6 \frac{\phi_0}{M_p} + \frac{2\dot\phi_0\phi_0}{HM_p^2}  \right),
\end{align}
where terms decaying as $a^{-3}$ or faster have been ignored. The last two terms decay as $a^{-3/2}$;
since resonances are induced by a linear combination of the two terms, let us first analyze the last term that is from back-reaction of the metric.

In this case, $\omega_k ^2\simeq k^2/a^2 + M^2 +2\dot\phi_0\phi_0 M^2/(HM_p^2)$.
Defining $\widetilde{\delta\phi}_k\equiv a^{3/2}\delta\phi_k$ and $\hat{T}\equiv Mt -\pi/4$, the perturbed Eq.~(\ref{eq_insHn0_5}) reads 
\begin{equation}
\label{eq_insHn0_7}
{\widetilde{\delta\phi}}_k''+\left[A_{3 k}-2q_{3}\cos(2 \hat{T})\right]\widetilde{\delta\phi}_k=0,
\end{equation}
where terms decaying as $a^{-3}$ (e.g. $H^2$ and $\dot{H}$) or faster have been ignored and 
\begin{eqnarray}
\label{eq_insHn0_8}
q_{3}&\equiv& \frac{a_{\rm ini}^3\phi_0^2(t_{\rm ini}) M  }{2a^3HM_p^2},\\
\label{eq_insHn0_9}
A_{3 k}&\equiv& 1+ \frac{k^2}{a^2 M^2}.
\end{eqnarray}
Since $q_3$ decreases below unity as $a^{-3}$, broad resonance does not happen. 
The narrow resonance condition~(\ref{eq:narrow}) yields for the first band
\begin{align}
 0 \le \frac{k}{M} \lesssim  a_{\rm ini}H_{\rm ini}\sqrt{ \frac{3a_{\rm ini}}{aHM} },
\end{align}
where we have used $3M_p^2H_{\rm ini}^2\simeq M^2\phi_0^2(t_{\rm ini})/2$. Thus, the first instability band starts from $k \lesssim a_{\rm ini}\sqrt{3H_{\rm ini}M}$ at $t = t_{\rm ini}$ and broadens as $a^{1/4}$.

Since the Floquet index is given by $\mu=q_3/2$, the modes in the first band grow as
\begin{align}
\label{eq_insHn0_9}
\widetilde{\delta\phi}_k \propto \exp{\left(\int \frac{q_3}{2} d\hat{T}\right)} 
&=\exp\left(\frac{ a_{\rm ini}^3\phi_0^2(t_{\rm ini})M^2}{2M_p^2}\int \frac{dt}{a^3H}\right)\nonumber\\
&\simeq \exp{\left( \frac32\int\frac{da}{a}\right)}.
\end{align}
Therefore, resonant modes of $\delta\phi_k$ stay constant as $\delta\phi_k=a^{-3/2}\widetilde{\delta\phi}_k =$ constant, which accounts for the low frequency regions of Figs.~\ref{FIG_6} and \ref{FIG_8}.

The second band appears within
\begin{align}
 3a^2 -   \frac{3 a_{\rm ini}^6H_{\rm ini}^4}{4 a^4H^2M^2} \lesssim \frac{k^2}{M^2} \lesssim  3a^2 + \frac{15 a_{\rm ini}^6H_{\rm ini}^4}{4 a^4H^2M^2},
\end{align}
whose width narrows as $a^{-1}$, and disappears quickly.

Let us next consider the second-to-last term in the frequency~(\ref{eq:omega_metric}) that is from self-interaction of the scalaron.
This case corresponds to subsection~\ref{subsec:friedmann}, and $\omega_k^2 \simeq k^2/a^2 + M^2 -\sqrt6 \phi_0 M^2/M_p$.
Defining $\widetilde{\delta\phi}_k\equiv a^{3/2}\delta\phi_k$ and $2\hat{T} \equiv Mt - \pi/2$, Eq.~(\ref{eq_insHn0_5}) reads Eq.~(\ref{eq_insHn0_7}) whose $q_3$ and $A_{3k}$ are replaced with
\begin{align}
q_4 &\equiv 2\sqrt6 \frac{\phi_0(t_{\rm ini})a_{\rm ini}^{3/2}}{M_p a^{3/2}},\\
A_{4k} &\equiv 4 + \frac{4k^2}{a^2M^2}.
\end{align} 
Although $q_4 = 4\sqrt6 > 1$ at $t=t_{\rm ini}$, the $q_4$-parameter gets smaller than unity within a few e-folds of the expansion as $q_4 \propto a^{-3/2}$. 
The first narrow resonance condition yields a band with
\begin{align}
\frac{k^2}{M^2} < -\frac34 a^2 +\frac{\sqrt6 \phi_0(t_{\rm ini})a_{\rm ini}^{3/2}\sqrt{a}}{2M_p},
\end{align}
which closes rapidly as $a^2$.
The second narrow resonance condition yields a band with
\begin{align}
\frac{k^2}{M^2} <  \frac{5\phi_0^2(t_{\rm ini})a_{\rm ini}^3}{2M_p^2a},
\end{align}
which narrows as $a^{-1}$. 
In Sec.~\ref{Sec_insH0}, we learned that it takes $t \gtrsim 10^3 \ [1/M]$ for instabilities to grow in the second resonant band; the value of $q_4$ is too small before this time.
In fact, we have confirmed that there is no instability in Fig.~\ref{FIG_5} by using a variable $a^{3/2}\delta\phi_k$ that counteracts the effect of the Hubble expansion. 
Thus, the resonance band closes and instabilities stop growing long before the growth overcomes the Hubble damping.

Finally, one may notice that there are intermediate regions of blue spectra with $|\delta\phi_k|^2 \propto k$ for $k\gtrsim M$ in Figs.~\ref{FIG_5} and \ref{FIG_6}. For sufficiently higher $k$-modes, the spectra are flat (not shown for $t = 10^3 \ [1/M]$ and $10^4 \ [1/M]$).
The appearance of the blue spectra can simply be understood by the time evolution of $\delta\phi_k$ in the absence of interactions. 
For $k > a M$, Eq.~(\ref{eq_insHn0_5}) is approximated by $\ddot{\delta\phi}_k +3H\dot{\delta\phi}_k +(k/a)^2\delta\phi_k \simeq 0$ whose solution is $\delta\phi_k \propto a^{-1}$; while for $k < aM$, $\ddot{\delta\phi}_k +3H\dot{\delta\phi}_k +M^2\delta\phi_k \simeq 0$ whose solution is $\delta\phi_k \propto a^{-3/2}$. Since physical $k$-modes are stretched by the cosmic expansion, more modes become ``heavy" and decay relatively faster as time passes. Thus, the flat spectra develop to the blue spectra for $ k < a M$.

\section{CONCLUSION}
\label{Sec_CONCLUSION}

In this paper, we have studied a possibility of the formation of soliton-like objects (oscillons/I-balls) after Starobinsky's $R^2$ inflation model. If they were formed, the reheating scenario would be significantly altered from the standard perturbative analysis, thereby resulting in modifications of the predictions for observable spectra of curvature and gravitational wave fluctuations.
Since the reheating after $R^2$ inflation proceeds through gravitational particle productions, the lifetime of the scalaron is rather long and oscillating excessively many times at the minimum of its potential~(\ref{eq:pot}).
In the meantime, the frequency $\omega_k$ of the scalaron fluctuations is periodically changing in time and parametric self-resonances occur in both narrow and broad instability bands, depending on the background spacetime.
If the scalaron fluctuations reach $\delta\phi \sim \phi_0$, soliton-like objects would be formed; thus, parametric amplification of the fluctuations is a necessary condition for oscillons/I-balls. More precisely, strong resonance with the Floquet index $[\mu_k M/H]_{\rm max} \gtrsim 10$ is a both necessary and sufficient condition for the formation of oscillons/I-balls (see the fourth Ref. of \cite{Amin:2010xe}).

In Sec.~\ref{Sec_insH0}, we have shown that parametric self-resonances induce instabilities in a static Minkowski background both numerically and analytically.
For the amplitude of the scalaron oscillations $\Phi < 0.2 M_p$, narrow resonance induces the exponential growth of fluctuation modes within the second instability band~(\ref{eq:second_narrow_minkowski}): $0 \le k < 2M\Phi/(\sqrt3 M_p)$. 
For $\Phi > 0.2 M_p$, broad resonance induces instabilities by violating the adiabaticity $|\dot\omega_k|/\omega_k^2 > 1$ for the modes with Eq.~(\ref{eq_sb_Anau_14}) if $\Phi \lesssim 2 M_p$ and for those with Eq.~(\ref{eq_sb_Anau_9}) if $\Phi \gtrsim 2 M_p$, respectively.
Therefore, the formation of oscillons/I-balls is likely to happen at particular scales in this case.

In Sec.~\ref{Sec_insHn0}, we have shown that parametric self-resonance is ineffective in an expanding Friedmann background. If the back-reaction from the metric is included, the metric preheating happens due to the first narrow resonance band with $0 \le k \lesssim  a_{\rm ini}H_{\rm ini}\sqrt{ 3a_{\rm ini}M/(aH) }$ in which the growth rate of scalaron fluctuations is balanced with the Hubble damping effect. In this case, however, resonance is not strong enough to form oscillons/I-balls since the Floquet index is too small.
Other resonance bands due to self-interactions disappear quickly as the universe expands.
Therefore, contrary to the Minkowski space analysis, the formation of oscillons/I-balls is not possible in an expanding universe.
As a result, the cosmological scenario in Sec.~\ref{Sec_model} is unchanged, holding the original predictions of the $R^2$ inflation.

The oscillating scalaron produces not only itself by the self-resonances but also almost all particles in the standard model as explained in Sec.~\ref{Sec_model}.
Although we have not included the back-reaction of the produced particles during preheating in Sec.~\ref{Sec_insHn0}, if this effect is taken into account, it leads to faster decay of the scalaron amplitude as $\langle\phi_0^2(t)\rangle\approx (\rho_{\phi, {\rm ini}}/M^2)(a/a_{\rm ini})^{-3}\exp{[-\Gamma_{\rm tot}(t-t_{\rm ini})]}$, where $ \rho_{\phi, {\rm ini}} \approx M^2 \phi_0^2(t_{\rm ini})/2$ and the scalaron decay rate $\Gamma_{\rm tot}$ is given by Eqs.~(\ref{eq:decay_scalar}), (\ref{eq:decay_spinor}) and (\ref{eq:decay_anomaly}).
Even though the factor $\exp{(-\Gamma_{\rm tot}t)}$ is not significant when $3H > \Gamma_{\rm tot}$, it further diminishes the chances of producing oscillons/I-balls.

Our result also applies to the symmetric potential (slightly shallower than quadratic away from the minimum) of conformal-type Higgs inflation~\cite{Bezrukov:2007ep} in the absence of massive gauge bosons.
It would be interesting to look for oscillons/I-balls formation in the Higgs inflation after which strong parametric resonances occur in the presence of electroweak gauge bosons~\cite{Bezrukov:2008ut}.

The above mentioned metric preheating was also reported in chaotic inflation models with monomial potentials~\cite{Nambu:1996gf,Finelli:1998bu,Jedamzik:2010dq, Easther:2010mr}.
Especially in Refs.~\cite{Easther:2010mr}, they pointed out the possibility of the breakdown of the coherent oscillations before the universe expands $\sim {\rm e}^{13}$ or so during the oscillation phase; the growth of density contrast $\delta\rho/\rho \sim 1$ makes the universe inhomogeneous.
In the $R^2$ inflation, the reheating temperature is predicted to be $\sim 10^9$ GeV; this instability is likely to grow to form non-linear small-scale structures, such as halos~\cite{Jedamzik:2010dq}, primordial black holes~\cite{Khlopov:1985jw}, and so on, which would also result in the gravitational wave signals detectable in the future space interferometer like DECIGO~(see the third Ref. of \cite{Bassett:1997ke} and the second Ref. of \cite{Gorbunov:2010bn}). 
If we extend gravitational interaction with matter sector to non-minimal one, preheating can be more efficient even in $R^2$ inflation~\cite{Tsujikawa:1999iv}.
We will leave these possibilities for future work.

\section*{Acknowledgments}
Y.W. thanks Mustafa Amin, Richard Easther, Teruaki Suyama and Jun'ichi Yokoyama for valuable discussions. Y.W. is grateful to Eiichiro Komatsu, Teruaki Suyama and Jun'ichi Yokoyama for useful comments on the earlier version of the draft.
The work of Y.W. was supported by Grant-in-Aids for Scientific Research on Innovative Areas No.~21111006, Scientific Research No.~23340058 and JSPS Fellow No.~269337.

\end{document}